\documentclass[12pt]{article}
\usepackage{epsfig}
\usepackage{color}
\usepackage{amssymb,amsmath}
\usepackage{graphicx}
\usepackage{here}
\def\nn{\nonumber}
\newcommand{\bea}{\begin{eqnarray}}
\newcommand{\eea}{\end{eqnarray}}
\newcommand{\mpt}{p\hspace{-0.45em}/ }

 \makeatletter
\def\alt{\mathrel{\mathpalette\gl@align<}}
\def\agt{\mathrel{\mathpalette\gl@align>}}
\def\gl@align#1#2{\lower.6ex\vbox{\baselineskip\z@skip\lineskip\z@
\ialign{$\m@th#1\hfil##\hfil$\crcr#2\crcr\sim\crcr}}} \makeatother

\begin{document}
\begin{flushright}
\end{flushright}
\vspace*{1.0cm}

\begin{center}
\baselineskip 20pt 
{\Large\bf 
Determining Strong Phase of $a_1$ Meson Decay Amplitude\\
using $W \to \nu \tau(\to \nu a_1(\to\pi^\mp \pi^\mp \pi^\pm))$ Process
}
\vspace{1cm}

{\large 
Kaoru Hagiwara$^a$, \ Hiroyuki Ishida$^a$, \ Toshifumi Yamada$^b$ and Daneng Yang$^c$
} \vspace{.5cm}

{\baselineskip 20pt \it
$^a$ KEK Theory Center, Tsukuba, Ibaraki 305-0801, Japan

$^b$ Institute of Science and Engineering, Shimane University, Matsue 690-8504, Japan

$^c$ Department of Physics, Tsinghua University, Beijing 100084, China
}

\vspace{.5cm}

\vspace{1.5cm} {\bf Abstract} \end{center}

To measure the helicity of a spin-1 meson from the triple vector product of the three-momenta of its decay products,
 one needs information about the strong phase of the decay amplitude.
In this paper, taking $a_1(1260)$ meson as an example, we present a method to
 extract information about the strong phase from the triple vector product of the pion momenta in
 $W \to \nu \tau(\to \nu a_1(\to\pi^\mp \pi^\mp \pi^\pm))$ process, where the $a_1$ helicity is known a priori from electroweak theory.
This process is advantageous in that highly-boosted $a_1^-$ mesons from $\tau_L^-$ decays
have nearly maximal helicity asymmetry and thus most reflect the strong phase.
We revisit the theoretical calculation of the $a_1$ meson helicity in $W \to \nu \tau(\to \nu a_1)$ process.
We formulate the differential decay rate of polarized $a_1$ mesons in a manner convenient for the study of the $a_1$ meson helicity asymmetry.
Finally, we present the method for extracting information about the strong phase, and assess its feasibility at the LHC.

\thispagestyle{empty}

\newpage

\baselineskip 18pt
%

\section{Introduction}

The helicity of spin-1 mesons can be a probe for physics beyond the Standard Model (SM).
For example, in $B^- \to K^- \pi^- \pi^+ \gamma$ process induced by $b\to s\gamma$,
 the SM predicts that the $K^- \pi^- \pi^+$ system is mostly left-handed
 because $W$ boson loop gives an amplitude with a left-handed photon,
 while various extensions of the SM contain an extra amplitude with a right-handed photon.
The helicity of the $K^- \pi^- \pi^+$ system can be determined from the triple vector product of the three-momenta of $K^-$, $\pi^-$, $\pi^+$,
 and indeed a non-zero polarization of the system has been confirmed experimentally~\cite{lhcb}.
Nevertheless, the helicity has not been measured.
The difficulty lies in the fact that the triple vector product of three-momenta is a na\"ive T-odd quantity~\cite{rujula}
 (odd under the reversal of all three-momenta and spins), and in CP-conserving theories like QCD
 its expectation value is non-zero only with the strong phase of the decay amplitude of $K^- \pi^- \pi^+$ resonances, which is poorly understood.
Since $K_1(1270)$ and $K_1(1400)$ resonances (the latter is much suppressed) 
contribute to $B^- \to K^- \pi^- \pi^+ \gamma$ process~\cite{Yang:2004as},
 efforts have been made to theoretically or phenomenologically determine the strong phase of $K_1(1270)$ and $K_1(1400)$ decay amplitudes~\cite{gronau1,gronau2,kou1,kou2,kou3,gronau3}.
Notably, Ref.~\cite{kou3} has pursued a purely phenomenological approach where one extracts,
 from experimental data on $B^- \to K^- \pi^- \pi^+ J/\psi$ process,
 information about the strong phase necessary for the $K_1(1270)$ helicity measurement.

\begin{figure}[H]
  \begin{center}
   \includegraphics[width=6cm]{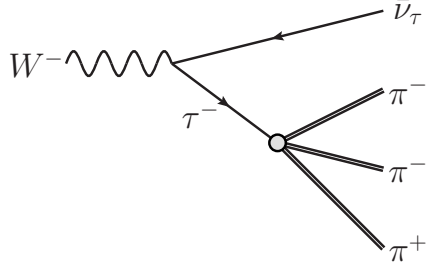}
 \caption{
 $W\to\nu\tau(\to \nu a_1(\to \pi^\mp \pi^\mp \pi^\pm))$ process.
 }
  \end{center}
  \label{taudecay}
\end{figure}

In this paper, we study experimental determination of the strong phase of a spin-1 meson's decay amplitude
 which utilizes a hadronic decay of $\tau$ lepton from a $W$ boson decay.
Since the helicity of a spin-1 meson in the decay of a polarized $\tau$ is known a priori from electroweak theory,
 we can use $W\to\nu\tau(\to \nu A)$ events ($A$ denotes a spin-1 meson) to determine the strong phase.
Moreover, $W\to\nu\tau(\to \nu A)$ process is advantageous in that
highly-boosted spin-1 mesons in $W\to\nu\tau(\to \nu A)$ events have nearly maximal helicity asymmetry (i.e. almost purely left-handed or right-handed) and hence the impact of the strong phase is maximized.
Although our ultimate target is the strong phase of $K_1(1270)$ and $K_1(1400)$ decay amplitudes, 
 we in this paper deal with a simpler case with $a_1(1260)$ meson.
We present a method to phenomenologically determine the strong phase of the $a_1^-\to\pi^-\pi^-\pi^+$ decay amplitude
\footnote{Throughout the paper, $a_1$ refers to $a_1(1260)$ meson.} 
\footnote{
Theoretical study on the hadronic form factors of $a_1^-\to\pi^-\pi^-\pi^+$ decay amplitude
 is found, e.g., in Refs.~\cite{Dumm:2009va,Shekhovtsova:2012ra,Nugent:2013hxa}.
}
from $W\to\nu\tau(\to \nu a_1(\to \pi^\mp \pi^\mp \pi^\pm))$ data (Fig.~\ref{taudecay}),
 and assess its feasibility in $W$ boson production events at the LHC.
Once the strong phase of the $a_1^-\to\pi^-\pi^-\pi^+$ decay amplitude is determined,
 one can use it to search for new physics through the $a_1$ polarization.
Moreover, we expect that the strong phase of the $K_1^-\to K^- \pi^- \pi^+$ decay amplitude is determined in basically the same manner,
 which is then utilized for the most interesting case, the photon polarization measurement in $B^- \to K^- \pi^- \pi^+ \gamma$ process.
\\

This paper is organized as follows:

In Section~2, we revisit the theoretical calculation of the helicity of $a_1^-$ meson in $W^- \to \bar{\nu}_\tau\tau^-(\to \nu_\tau a_1^-)$ process.
The $a_1^-$ helicity is calculated as a function of the energy fraction of $a_1^-$ in $\tau^-$ decay in the laboratory frame, $z=E_{a_1}/E_\tau$.
We will confirm that $a_1^-$ with $z\gtrsim0.8$ (i.e. highly-boosted $a_1^-$) is almost purely left-handed.

In Section~3, we express the differential decay rate of polarized $a_1^-$ mesons using the following parametrization:
Let $p_1,p_2,p_3$ respectively denote the four-momenta of $\pi^-,\pi^-,\pi^+$, with $Q\cdot p_1 > Q\cdot p_2$
 ($Q=p_1+p_2+p_3$).
In an $a_1^-$ rest frame, we write the angle between $\vec{p}_3 \times \vec{p}_1$ and $a_1^-$'s boost direction in the laboratory frame
 as $\Psi$, and write the angle between $\vec{p}_3$ and the projection of $a_1^-$'s boost direction onto the $a_1^-$ decay plane as $\phi$.
The angles $\Psi,\,\phi$ and the Dalitz variables $s_{13}=(p_1+p_3)^2,\,s_{23}=(p_2+p_3)^2$ 
 completely parametrize the differential decay rate.
A benefit of this parametrization is that that part of the differential decay rate which reflects the $a_1^-$ helicity asymmetry is simply linear in $\cos\Psi$ and is independent of $\phi$.
\footnote{
K\"uhn-Mirkes parametrization~\cite{ks} has been widely used for describing the $\tau^- \to 3\pi\nu_\tau$ decay kinematics.
Our parametrization differs from it in that the coordinate is defined independently of $\tau^-$ momentum, namely,
 purely the $a_1^- \to 3\pi$ decay kinematics is described.
}

In Section~4, we present a method to determine the strong phase using $W\to\nu\tau(\to \nu a_1(\to \pi^\mp \pi^\mp \pi^\pm))$ events,
 based on the theoretical calculation of the $a_1^-$ helicity in Section~2 and the parameterization of the differential decay rate in Section~3.
Statistical uncertainty in the above determination at the 14~TeV LHC with 300~fb$^{-1}$ of data is further estimated.

Section~5 summarizes the paper.

In Appendix, we give a simple derivation of the Wigner rotation, which is used in the calculation of the $a_1^-$ helicity in boosted $\tau^-$ decays
 in Section~2.
\\

\section{$a_1^-$ Helicity in $W^-\to\bar{\nu}_{\tau}\tau^-(\to\nu_\tau a_1^-)$ Process}

Since $\tau^-$ in $W^-\to \tau^- \bar{\nu}_\tau$ process is almost purely left-handed,
 it suffices to consider polarized $\tau^-$.
The differential decay rate of $\tau^- \to \nu_{\tau} \pi^- \pi^- \pi^+$ process with polarized $\tau^-$ is expressed as
\begin{align}
&{\rm d}\Gamma(\tau_h^- \to \pi^- \pi^- \pi^+ \nu_{\tau}) \nonumber \\
&= \dfrac{1}{2 m_{\tau}} \left| \sum_{\lambda=\pm,0} \, {\cal M}(\tau_h^- \to \nu_{\tau}\,a_{1,\lambda}^-) \, B_{a_1}(Q^2) \, {\cal M}(a_{1,\lambda}^- \to \pi^- \pi^- \pi^+) \right|^2 {\rm d}\Phi_2(\tau \to a_1 \nu) \frac{{\rm d}Q^2}{2\pi} {\rm d}\Phi_3(a_1 \to 3\pi),
\label{diffrate}\\
&\ \ \ \ \ \ \ h=\pm\frac{1}{2}:\tau^- \ {\rm helicity},\ \ \ \ \ \lambda=\pm,0:a_1^- \ {\rm helicity}, \ \ \ \ \ Q^\mu:a_1^- \ {\rm momentum}.
\nn
\end{align}
${\cal M}(\tau_h^- \to \nu_{\tau}\,a_{1,\lambda}^-)$ denotes the helicity amplitude of $\tau^- \to \nu_{\tau}a_1^-$ process, 
 $B_{a_1}(Q^2)$ is the form factor of $a_1$ resonance that satisfies $B(Q^2=0)=1$, and ${\cal M}(a_{1,\lambda}^- \to \pi^- \pi^- \pi^+)$ denotes the helicity amplitude of $a_1^- \to \pi^- \pi^- \pi^+$ process.
d$\Phi_2(\tau\to a_1\nu)$ and d$\Phi_3(a_1\to3\pi)$ denote the phase space factors of $\tau^- \to \nu_{\tau} a_1^-$ and $a_1^- \to \pi^- \pi^- \pi^+$ processes, respectively.
In Eq.~(\ref{diffrate}), the contribution from $\pi(1300)$ resonance is neglected,
 since the OPAL Collaboration has reported, based on a fitting of $\tau^-\to\pi^-\pi^-\pi^+\nu$ data, that 
 the branching ratio of $\tau^- \to \pi^-(1300)\nu$ process is below 0.84\% of the total $\tau^-\to\pi^-\pi^-\pi^+\nu$ branching ratio~\cite{Ackerstaff:1997dv}.
Also, a single axial-vector resonance, $a_1(1260)$, is assumed to dominate the process,
 since the CLEO Collaboration has reported that the same assumption yields a good fit to isospin-related process $\tau^-\to\pi^-\pi^0\pi^0\nu$~\cite{Asner:1999kj}.
Note that the vector current contribution is negligible in $\tau^- \to \pi^-\pi^-\pi^+\nu_\tau$ process
 due to $G$-parity of QCD.

Integrating out the azimuthal angle of $a_1^-$ momentum around the $\tau^-$ helicity axis,
we remove interference among amplitudes with different $a_1^-$ helicities.
The differential decay rate is then factorized into the one for $\tau^-\to \nu_\tau a_1^-$ process and the one for $a_1^-\to \pi^-\pi^-\pi^+$ process,
 and is expressed as
\begin{align}
{\rm d}\Gamma(\tau_h^- \to \nu_{\tau} \pi^- \pi^- \pi^+)
&=\dfrac{1}{2 m_{\tau}} \sum_{\lambda=\pm1,0} \, \left| {\cal M}(\tau_h^- \to \nu_{\tau}\,a_{1,\lambda}^-) \right|^2 \, |B_{a_1}(Q^2)|^2 \, \left| {\cal M}(a_{1,\lambda}^- \to \pi^- \pi^- \pi^+) \right|^2 \nonumber \\
&\ \ \ \ \ \ \ \ \ \ \ \ \ \ \times \frac{1}{16\pi}\left( 1 - \frac{Q^2}{m_{\tau}^2} \right) {\rm d}\cos\hat{\theta} \frac{{\rm d}Q^2}{2\pi} {\rm d}\Phi_3(a_1 \to 3\pi)
\label{diffratelabpre}
\end{align}
 where $\hat{\theta}$ denotes the angle between the $a_1^-$ momentum and the $\tau^-$ helicity axis in a $\tau^-$ rest frame.
For convenience, we trade $\hat{\theta}$ for the energy fraction of $a_1^-$ in $\tau^-$ decay in the laboratory frame, $z$,
\begin{align}
z &= \frac{E_{a_1}}{E_{\tau}} = \frac{1+Q^2/m_\tau^2+\beta(1-Q^2/m_\tau^2)\cos\hat{\theta}}{2}
\end{align}
 where $\beta$ denotes the speed of $\tau^-$ in the laboratory frame.
Eq.~(\ref{diffratelabpre}) is then rewritten as
\begin{align}
 {\rm d}\Gamma(\tau_h^- \to \nu_{\tau} \pi^- \pi^- \pi^+)&=|B_{a_1}(Q^2)|^2 2\sqrt{Q^2} \frac{{\rm d}Q^2}{2\pi}
\sum_{\lambda=\pm1,0}\frac{{\rm d}\Gamma(\tau_h^- \to \nu_{\tau}\,a_{1,\lambda}^-)}{{\rm d}z} {\rm d}z \,
{\rm d}\Gamma(a_{1,\lambda}^- \to \pi^- \pi^- \pi^+),
\label{diffratelab}
\end{align}
 with
\begin{align} 
\frac{{\rm d}\Gamma(\tau_h^- \to \nu_{\tau}\,a_{1,\lambda}^-)}{{\rm d}z}&=
\frac{1}{\beta}\frac{1}{16\pi m_\tau}\left| {\cal M}(\tau_h^- \to \nu_{\tau}\,a_{1,\lambda}^-) \right|^2,
\label{diffratetau}\\
{\rm d}\Gamma(a_{1,\lambda}^- \to \pi^-\pi^-\pi^+)&=\frac{1}{2\sqrt{Q^2}}\left| {\cal M}(a_{1,\lambda}^- \to \pi^- \pi^- \pi^+) \right|^2{\rm d}\Phi_3(a_1 \to 3\pi).
\label{diffratea1}
\end{align}
${\rm d}\Gamma(\tau_h^- \to \nu_{\tau}\,a_{1,\lambda}^-)/{\rm d}z$ corresponds to the differential decay rate of $\tau^-$ with helicity $h$
 decaying into $a_1^-$ with helicity $\lambda$, for a specific value of $z$.
d$\Gamma(a_{1,\lambda}^- \to \pi^-\pi^-\pi^+)$ corresponds to the differential decay rate of $a_1^-$ with helicity $\lambda$.
${\rm d}\Gamma(\tau_h^- \to \nu_{\tau}\,a_{1,\lambda}^-)/{\rm d}z$ for $h=-1/2$ encodes
 the $a_1^-$ helicity distribution in $W^-\to\bar{\nu}_{\tau}\tau^-(\to\nu_\tau a_1^-)$ process.
\\

In the rest of the section, we evaluate ${\rm d}\Gamma(\tau_h^- \to \nu_{\tau}\,a_{1,\lambda}^-)/{\rm d}z$.
The helicity amplitude ${\cal M}(\tau_h^- \to \nu_{\tau}\,a_{1,\lambda}^-)$ is given by
\begin{align}
{\cal M}(\tau_h^- \to \nu_{\tau}\,a_{1,\lambda}^-) &= \sqrt{2}G_F\cos\theta_C \ \bar{\nu}_{\tau}\gamma^{\mu} \frac{1-\gamma_5}{2}\tau\left(\vec{p}_{\tau},h\right) \epsilon_{\mu}^*\left(Q^2,\lambda\right),
\label{taudecayamp}
\end{align}
 where we retain $Q^2$ dependence of the polarization vector $\epsilon_{\mu}$, since $a_1$ is a broad resonance.
The helicity amplitude ${\cal M}(\tau_h^- \to \nu_{\tau}\,a_{1,\lambda}^-)$ is specified
 in terms of the $a_1^-$ helicity along the $a_1^-$ boost direction in a $\tau^-$ rest frame, $\lambda_\tau$,
 and the angle between the $a_1^-$ momentum and the $\tau^-$ helicity axis in a $\tau^-$ rest frame $\hat{\theta}$; we find
\begin{align}
{\cal M}(\tau_h^- \to \nu_{\tau}\,a_{1,\lambda_{\tau}}^-) &= \sqrt{2}G_F\cos\theta_C \sqrt{m_{\tau}^2-Q^2} \, {\cal \hat{M}}_{h \lambda_\tau},\label{taudecayamp2}
\end{align}
 where
\begin{align}
{\cal \hat{M}}_{-\frac{1}{2},-} &=  \sqrt{2} \cos\frac{\hat{\theta}}{2}, \ \ \ \ \ {\cal \hat{M}}_{-\frac{1}{2},0} = \frac{m_\tau}{\sqrt{Q^2}} \sin\frac{\hat{\theta}}{2}, \ \ \ \ \ {\cal \hat{M}}_{-\frac{1}{2},+} = 0, \\
{\cal \hat{M}}_{\frac{1}{2},-} &= -\sqrt{2} \sin\frac{\hat{\theta}}{2}, \ \ \ \ \ {\cal \hat{M}}_{\frac{1}{2},0} = \frac{m_\tau}{\sqrt{Q^2}}\cos\frac{\hat{\theta}}{2}, \ \ \ \ \ {\cal \hat{M}}_{\frac{1}{2},+} = 0.
\end{align}

Experimentally, what we measure is the $a_1^-$ helicity along the $a_1^-$ boost direction in the laboratory frame, $\lambda_{\rm lab}$, 
 not the helicity along the $a_1^-$ boost direction in a $\tau^-$ rest frame $\lambda_\tau$.
Hence, we want to rewrite Eq.~(\ref{taudecayamp2}) in terms of $\lambda_{\rm lab}$.
For this purpose, we expand the $a_1^-$ polarization vectors in the laboratory frame in terms of those in a $\tau$ rest frame, as
\begin{align}
\epsilon^{\mu}(Q^2, \, \lambda_{\rm lab}) &= \sum_{\lambda_\tau=\pm1,0} \left\{ -\epsilon^{\mu}(Q^2, \, \lambda_\tau)\epsilon^{\nu *}(Q^2, \, \lambda_\tau) \right\} \epsilon_{\nu}(Q^2, \, \lambda_{\rm lab})\nonumber\\
&= \sum_{\lambda_\tau=\pm1,0} \left\{ -\epsilon^{\nu *}(Q^2, \, \lambda_\tau)\epsilon_{\nu}(Q^2, \, \lambda_{\rm lab}) \right\} \epsilon^{\mu}(Q^2, \, \lambda_\tau) \nonumber \\
&= \sum_{\lambda_\tau=\pm1,0} d^{J=1}_{\lambda_\tau\lambda_{\rm lab}}(\tilde{\theta}) \epsilon^{\mu}(Q^2, \, \lambda_\tau),
\end{align}
 where $d_{\lambda' \lambda}^{J=1}$ is a $d$-function, and
$\tilde{\theta}$ is the angle between the $a_1^-$ boost direction in a $\tau^-$ rest frame and that in the laboratory frame, measured in an $a_1^-$ rest frame.
$\tilde{\theta}$ is expressed in terms of the angle between the $a_1^-$ momentum and the $\tau^-$ helicity axis in a $\tau^-$ rest frame $\hat{\theta}$,
 and the speed and boost factor of $\tau^-$ in the laboratory frame $\beta$ and $\gamma=1/\sqrt{1-\beta^2}$, as (see Appendix for the derivation)
\begin{align}
\cos \tilde{\theta} &= \frac{(1+a^2) \beta \cos\hat{\theta} + 1 - a^2}{\sqrt{ \left\{ (1+a^2)+(1-a^2)\beta \cos\hat{\theta} \right\}^2 - 4a^2/\gamma^2 }}
\ \ \ \ \ \ (a = \sqrt{Q^2}/m_{\tau}),
\label{wignerangle}
\end{align}
 or equivalently, in terms of the energy fraction of $a_1^-$ in $\tau^-$ decay in the laboratory frame $z$, as
\begin{align}
\cos\tilde{\theta} &= \frac{z(1+a^2)-2a^2}{(1-a^2)\sqrt{ z^2-\frac{a^2}{\gamma^2} }} \ \ \ \ \ \ \ \ (z=E_{a_1}/E_\tau).
\label{wignerangle-alt}
\end{align}
The helicity amplitude is rewritten in terms of the $a_1^-$ helicity in the laboratory frame $\lambda_{\rm lab}$, as
\begin{align}
{\cal M}(\tau_h^- \to \nu_{\tau}\,a_{1,\lambda_{\rm lab}}^-) &= \sum_{\lambda_\tau=\pm1,0} d^{J=1}_{\lambda_\tau\lambda_{\rm lab}}(\tilde{\theta}) {\cal \hat{M}}_{h \lambda_\tau}(\hat{\theta})
\label{wignerrotation}
\end{align}
 with $\tilde{\theta}$ given in Eq.~(\ref{wignerangle}) or Eq.~(\ref{wignerangle-alt}).

\begin{figure}[H]
  \begin{center}
    \includegraphics[width=8cm]{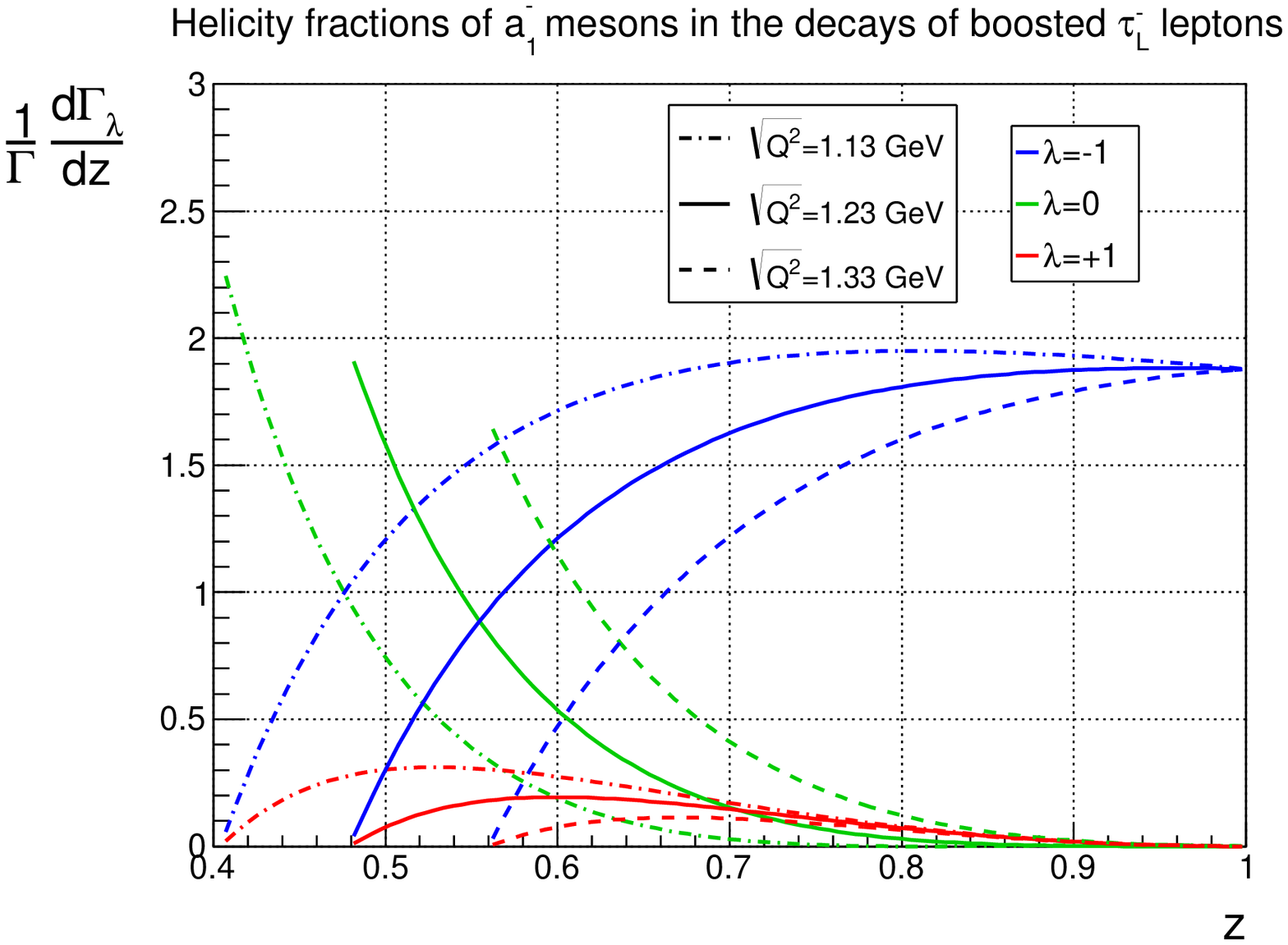}
    \includegraphics[width=8cm]{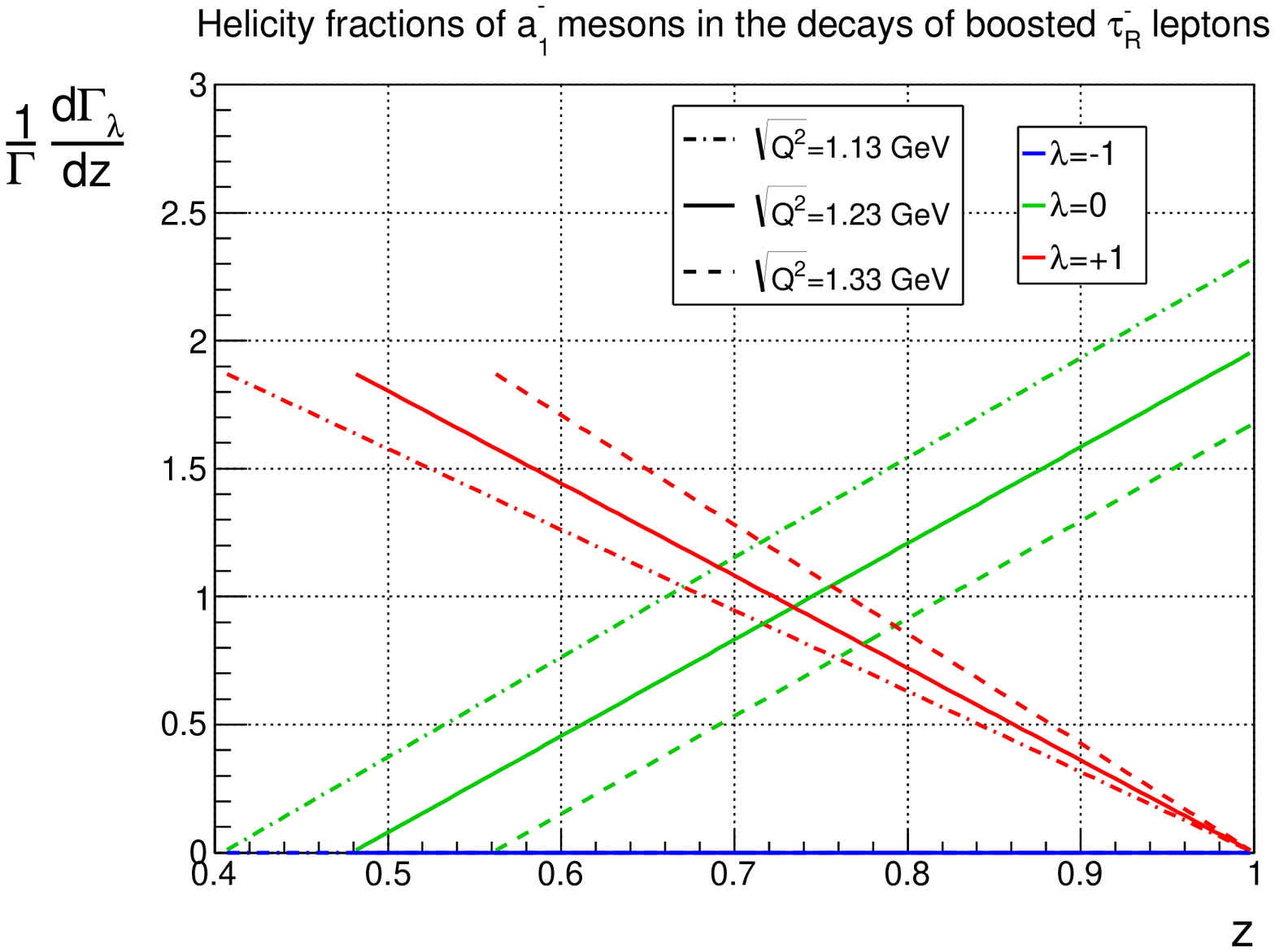}
    \caption{Left: Normalized differential decay rate Eq.~(\ref{normalizeddiffrate}) for $\tau^-$ with helicity $h=-1/2 \ (\tau_L^-)$,
      for each final-state $a_1^-$ helicity in the laboratory frame $\lambda_{\rm lab}=\pm,0$.
    The horizontal axis is $z=E_{a_1}/E_\tau$, the energy fraction of $a_1^-$ in $\tau^-$ decay in the laboratory frame.
    The boost factor of $\tau^-$ in the laboratory frame is fixed as $\gamma=23$.
    The dot-dashed, solid and dashed lines correspond to different $a_1^-$ invariant masses $\sqrt{Q^2}=1.13$~GeV, 1.23~GeV, 1.33~GeV.
    Right: The same as the left plot except that $\tau^-$ has helicity $h=1/2 \ (\tau_R^-)$.
    }
    \label{fracLR}
  \end{center}
\end{figure}

Assembling Eqs.~(\ref{diffratetau}),(\ref{taudecayamp2}),(\ref{wignerrotation}),(\ref{wignerangle-alt}), we numerically calculate 
 ${\rm d}\Gamma(\tau_h^- \to \nu_{\tau}\,a_{1,\lambda}^-)/{\rm d}z$
 and present it in the form of the normalized differential decay rate
\begin{align}
\frac{1}{\Gamma} \frac{{\rm d}\Gamma_{\lambda_{\rm lab}}}{{\rm d}z}(z,Q^2) &=
\frac{1}{\sum_{\lambda=\pm1,0}\int{\rm d}z'
\frac{{\rm d}\Gamma(\tau_h^- \to \nu_{\tau}\,a_{1,\lambda}^-;\,z',Q^2)}{{\rm d}z'}
}\frac{{\rm d}\Gamma(\tau_h^- \to \nu_{\tau}\,a_{1,\lambda_{\rm lab}}^-;\,z,Q^2)}{{\rm d}z}
\label{normalizeddiffrate}
\end{align}
 in Fig.~\ref{fracLR},
 for each $\tau^-$ helicity $h=-1/2$ ($\tau_L^-$) and $h=1/2$ ($\tau_R^-$).
The boost factor of $\tau^-$ is fixed as $\gamma=23$, corresponding to the boost factor of $\tau^-$ from the decay of a $W$ boson at rest.
However, the plots are almost independent of $\gamma$ when $\gamma \gtrsim 3$.
Three different values of the $a_1^-$ invariant mass, $\sqrt{Q^2}=1.13$~GeV, 1.23~GeV, 1.33~GeV, are considered.

\noindent
The left panel of Fig.~\ref{fracLR} presents the $a_1^-$ helicity in $W^-\to\bar{\nu}_{\tau}\tau^-(\to\nu_\tau a_1^-)$ process.
In the left panel, we observe that $a_1^-$ meson with $z\gtrsim0.8$ is almost purely left-handed ($\lambda_{\rm lab}=-1$).

We note that Fig.~\ref{fracLR} is in agreement with the preceding study Ref.~\cite{bhm}.
\\

\section{Differential Decay Rate of Polarized $a_1^-$ Mesons}

We formulate the differential decay rate of $a_1^- \to \pi^- \pi^- \pi^+$ process for each $a_1^-$ helicity, Eq.~(\ref{diffratea1}),
 in a manner convenient for the study of the helicity asymmetry.

The helicity amplitude is written as
\begin{align}
{\cal M}(a_{1,\lambda}^- \to \pi^- \pi^- \pi^+)&= \epsilon^{\mu}(Q^2,\lambda) J_{\mu},
\label{a1decayamp}
\end{align}
 where $\epsilon^{\mu}(Q,\lambda)$ is the polarization vector of $a_1^-$ with helicity $\lambda$, and
 $J_{\mu}$ is the hadronic current,
\begin{align} 
J^\mu&= \langle \pi^- \pi^- \pi^+\vert (-\bar{u}\gamma^\mu\gamma^5d)\vert 0\rangle.
\label{jmu}
\end{align}
The most general parametrization for the hadronic current $J^\mu$ that respects
 (i) Lorentz covariance, (ii) the current conservation $Q^\mu J_\mu=0$, and (iii) Bose symmetry of two $\pi^-$'s, is given as follows:
Let $p_1,p_2,p_3$ respectively denote the momenta of $\pi^-,\pi^-,\pi^+$ (then $Q^\mu=(p_1+p_2+p_3)^\mu$), where the two $\pi^-$'s are distinguished by $Q\cdot p_1>Q\cdot p_2$.
The most general parametrization is then
\begin{align} 
J^{\mu}(p_1,p_2,p_3)=&-i\frac{2\sqrt{2}}{3f_\pi} \left\{ \, ( p_{1\nu} - p_{3\nu} ) F(Q^2,s_{13},s_{23}) + ( p_{2\nu} - p_{3\nu} ) F(Q^2,s_{23},s_{13}) \, \right\} \left( g^{\mu\nu} - \frac{Q^{\mu}Q^{\nu}}{Q^2} \right),
\label{generalhadcurrent}
\end{align}
 where $s_{13}\equiv(p_1+p_3)^2$ and $s_{23}\equiv(p_2+p_3)^2$, $f_\pi\simeq93$~MeV is the pion decay constant,
 and $F(Q^2,s_{13},s_{23})$ is a general function of three Lorentz scalars.
$F(Q^2,s_{13},s_{23})$ is normalized in such a way that
if $\pi\pi$ resonances were absent, we would have $F(Q^2,s_{13},s_{23})=1$ by chiral perturbation theory~\cite{Fischer:1979fh}.

We explicitly write the $a_1^-$ decay helicity amplitude Eq.~(\ref{a1decayamp})
 in the $a_1^-$ rest frame whose $z$-axis is along the $a_1^-$ boost direction in the laboratory frame
 (thus $\lambda_{\rm lab}$ is along this $z$-axis).
In this frame, the momenta of the three pions and their sum can be parametrized as
\begin{align}
Q^{\mu} &= (\sqrt{Q^2}, \, 0, \, 0, \, 0),
\\
p_3^{\mu} &= (E_3, \, \vec{p}_3) = \frac{1}{2\sqrt{Q^2}}
\left( \begin{array}{c} x_3 \\ \tilde{x}_3 \cos\phi \cos\Psi \\ \tilde{x}_3 \sin\phi \\ -\tilde{x}_3 \sin\Psi\cos\phi \end{array} \right), 
\label{p3momentum}\\
p_1^{\mu} &= (E_1, \, \vec{p}_1) = \frac{1}{2\sqrt{Q^2}}
\left( \begin{array}{c} x_1 \\
 \tilde{x}_1 \cos\Psi ( \cos\phi\cos\theta_1 - \sin\phi\sin\theta_1 ) \\
 \tilde{x}_1 ( \sin\phi\cos\theta_1 + \cos\phi\sin\theta_1 ) \\
 -\tilde{x}_1 \sin\Psi ( \cos\phi\cos\theta_1 - \sin\phi\sin\theta_1 ) \end{array} \right), 
\label{p1momentum}\\ 
p_2^{\mu} &= (E_2, \, \vec{p}_2) = \frac{1}{2\sqrt{Q^2}} 
\left( \begin{array}{c} x_2 \\
 \tilde{x}_2 \cos\Psi ( \cos\phi\cos\theta_2 + \sin\phi\sin\theta_2 ) \\
 \tilde{x}_2 ( \sin\phi\cos\theta_2 - \cos\phi\sin\theta_2 ) \\
 -\tilde{x}_2 \sin\Psi ( \cos\phi\cos\theta_2 + \sin\phi\sin\theta_2 ) \end{array} \right),
 \label{p2momentum}
\end{align}
 where $x_i$ and $\tilde{x}_i$ are defined in terms of $Q^2,s_{13},s_{23}$ as
\begin{align} 
x_i &= 2Q\cdot p_i = Q^2-s_{jk}+m_\pi^2 \ \ \ \ \  ((i,j,k)=(1,2,3),(2,3,1),(3,1,2)),
\\
\tilde{x}_i &= \sqrt{x_i^2 - 4 Q^2m_\pi^2}.
\label{x-def}
\end{align}
$\theta_i$ ($i=1,2$) denotes the angle between $\vec{p}_i$ and $\vec{p}_3$, which is given in terms of $\tilde{x}_i$ as
\begin{align} 
\cos\theta_1 &= \frac{\tilde{x}_2^2 - \tilde{x}_1^2 - \tilde{x}_3^2}{2 \tilde{x}_1 \tilde{x}_3},\ \ \ \ \
\cos\theta_2 = \frac{\tilde{x}_1^2 - \tilde{x}_2^2 - \tilde{x}_3^2}{2 \tilde{x}_2 \tilde{x}_3}.
\end{align}
$\Psi$ is the angle between $\vec{p}_3 \times \vec{p}_1$ vector and the $z$-axis,
  and $\phi$ is the angle between $\vec{p}_3$ vector and the projection of the $z$-axis onto the $a_1^-$ decay plane,
  which satisfy
\begin{align}
\vec{p}_3 \times \vec{p}_1 &= \frac{1}{4Q^2}\tilde{x}_3\tilde{x}_1\sin\theta_1 (\sin\Psi, \, 0, \, \cos\Psi).
\end{align}
The polarization vectors are given by
\begin{align}
\epsilon^{\mu}(Q^2, \, \lambda_{\rm lab}=\pm) = \frac{1}{\sqrt{2}}(0, \, \mp1, \, -i, \, 0), \ \ \ \ \ \epsilon^{\mu}(Q^2, \, \lambda_{\rm lab}=0) = (0, \, 0, \, 0, \, 1).
\label{a1pol}
\end{align}
From Eqs.~(\ref{a1decayamp}),(\ref{generalhadcurrent}),(\ref{a1pol}), the helicity amplitudes are expressed as
\begin{align}
{\cal M}(a_{1,\lambda_{\rm lab}=\pm}^- \to \pi^- \pi^- \pi^+) = &\frac{1}{3 f_{\pi}} \frac{1}{\sqrt{Q^2}} \left[ \pm\cos\Psi \left\{ \cos\phi \ A(Q^2,s_{13},s_{23}) - \sin\phi \ B(Q^2,s_{13},s_{23}) \right\}\right.
\nonumber\\
&\left. \ \ \ \ \ \ \ \ \ \ \ \ \ \ \ \ \ \ \ \ + i \left\{ \sin\phi \ A(Q^2,s_{13},s_{23}) + \cos\phi \ B(Q^2,s_{13},s_{23}) \right\}  \right],
\label{amp+}\\
{\cal M}(a_{1,\lambda_{\rm lab}=0}^- \to \pi^- \pi^- \pi^+) = &\frac{\sqrt{2}}{3 f_{\pi}} \frac{1}{\sqrt{Q^2}} \sin\Psi \left\{ \cos\phi \ A(Q^2,s_{13},s_{23}) - \sin\phi \ B(Q^2,s_{13},s_{23}) \right\},
\label{amp0}
\end{align}
 where $A$ and $B$ are structure functions with mass dimension $+2$ defined as
  \footnote{
In the $m_\pi \to 0$ limit, they asymptote as
\begin{align}
A(Q^2,s_{13},s_{23}) &\xrightarrow{m_{\pi}\to 0} \ \left( x_1 - x_3 - \dfrac{2(1-x_2)}{x_3} \right) F(Q^2,s_{13},s_{23}) + \left(x_2-x_3-\dfrac{2(1-x_1)}{x_3}\right) F(Q^2,s_{23},s_{13}), 
\label{Alimit}\\
B(Q^2,s_{13},s_{23}) &\xrightarrow{m_{\pi}\to 0} \ \dfrac{2}{x_3} \sqrt{(1-x_1)(1-x_2)(1-x_3)} \left\{ F(Q^2,s_{13},s_{23}) - F(Q^2,s_{23},s_{13}) \right\}.
\label{Blimit}
\end{align}
}
(remind that $\tilde{x}_i$ is related to $Q^2,s_{13},s_{23}$ through Eq.~(\ref{x-def}))
\begin{align}
A(Q^2,s_{13},s_{23}) &= (\cos\theta_1 \tilde{x}_1-\tilde{x}_3) F(Q^2,s_{13},s_{23}) + (\cos\theta_2 \tilde{x}_2-\tilde{x}_3) F(Q^2,s_{23},s_{13}),  \label{Adef} \\
B(Q^2,s_{13},s_{23}) &= \sin\theta_1 \tilde{x}_1 \left\{ F(Q^2,s_{13},s_{23}) - F(Q^2,s_{23},s_{13}) \right\}.
\label{Bdef}
\end{align}

Finally, we plug Eqs.~(\ref{amp+}),(\ref{amp0}) into the formula for the polarized $a_1^-$ differential decay rate
\begin{align} 
{\rm d}\Gamma(a_{1,\lambda_{\rm lab}}^- \to \pi^-\pi^-\pi^+)
 &=\frac{1}{2\sqrt{Q^2}}\vert {\cal M}(a_{1,\lambda_{\rm lab}}^- \to \pi^- \pi^- \pi^+) \vert^2 {\rm d}\Phi_3(a_1 \to 3\pi)
\nn\\
&=\frac{1}{2\sqrt{Q^2}}\vert {\cal M}(a_{1,\lambda_{\rm lab}}^- \to \pi^- \pi^- \pi^+) \vert^2
\dfrac{1}{128 \pi^3} \dfrac{1}{Q^2} {\rm d}s_{13} {\rm d}s_{23} \dfrac{{\rm d}\cos\Psi}{2} \dfrac{{\rm d}\phi}{2 \pi}
\nn
\end{align}
 and obtain
\begin{align}
&\frac{{\rm d}^4\Gamma(a_{1,\lambda_{\rm lab}=\pm}^-\to\pi^-\pi^-\pi^+)}{{\rm d}\cos\Psi {\rm d}\phi {\rm d}s_{13} {\rm d}s_{23}} \nonumber \\
 &= \dfrac{1}{512 \pi^4} \frac{1}{2Q^4\sqrt{Q^2}} \dfrac{1}{9 f_{\pi}^2} \,
 \left[ \, |A|^2 + |B|^2 - (1-\cos^2\Psi)\left\{ \, \cos^2\phi |A|^2 + \sin^2\phi |B|^2 - \sin2\phi \, {\rm Re}(A\cdot B^*) \, \right\} \right.
\nonumber \\
 &\left. \ \ \ \ \ \ \ \ \ \ \ \ \ \ \ \ \ \ \ \ \ \ \ \ \ \ \ \ \ \ \ \ \ \ \ \ \ \ \ \ \ \ \ \ \ \ \ \ \ \ \ \ \ \ \ \ \ \ \ \ \ \ \ \ \ \ \ \ \ \ \ \ \ \ \ \ \ \ \ \ \ \ \ \ \ \
 \pm 2\cos\Psi \, {\rm Im}(A\cdot B^*) \, \right], 
\label{diffrate+} \\
&\frac{{\rm d}^4\Gamma(a_{1,\lambda_{\rm lab}=0}^-\to\pi^-\pi^-\pi^+)}{{\rm d}\cos\Psi {\rm d}\phi {\rm d}s_{13} {\rm d}s_{23}} \nonumber \\
 &= \dfrac{1}{512 \pi^4} \frac{1}{2Q^4\sqrt{Q^2}} \dfrac{1}{9 f_{\pi}^2} \, 
 2(1-\cos^2\Psi)\left\{ \, \cos^2\phi |A|^2 + \sin^2\phi |B|^2 - \sin2\phi \, {\rm Re}(A\cdot B^*) \, \right\}.
\label{diffrate0}
\end{align}
\\

Consider a general $a_1$ production process, not limited to the $W\to\nu\tau(\to a_1 \nu)$ process.
In terms of the transverse and asymmetric helicity fractions in the laboratory frame defined by
\begin{align}
P_T &=\frac{ ({\rm Number \ of \ } a_1^- {\rm \ mesons \ with \ } \lambda_{\rm lab}=+) + ({\rm Number \ of \ } a_1^- {\rm \ mesons \ with \ } \lambda_{\rm lab}=-) }{({\rm Total \ number \ of \ } a_1^- {\rm \ mesons})},
\nn\\
P_A &=\frac{ ({\rm Number \ of \ } a_1^- {\rm \ mesons \ with \ } \lambda_{\rm lab}=+) - ({\rm Number \ of \ } a_1^- {\rm \ mesons \ with \ } \lambda_{\rm lab}=-) }{({\rm Total \ number \ of \ } a_1^- {\rm \ mesons})},
\nn
\end{align}
 the $a_1^-$ differential decay rate in a general $a_1$ production process satisfies
\begin{align}
&\frac{1}{\Gamma(a_1^-\to\pi^-\pi^-\pi^+)}\frac{{\rm d}^4\Gamma(a_1^-\to\pi^-\pi^-\pi^+)}{{\rm d}\cos\Psi {\rm d}\phi {\rm d}s_{13} {\rm d}s_{23}}\nonumber\\
&=P_T \frac{1}{4\pi N_{{\rm nor}}} \left[|A|^2 + |B|^2 - 3 (1-\cos^2\Psi) \left\{\cos^2\phi |A|^2 + \sin^2\phi |B|^2 - \sin2\phi \, {\rm Re}(A\cdot B^*)\right\}\right]
\label{pdfT}\\
&+P_A \frac{1}{4\pi N_{{\rm nor}}} 2\cos\Psi \ {\rm Im}(A\cdot B^*)
\label{pdfA} \\
&+\frac{1}{4\pi N_{{\rm nor}}} 2(1-\cos^2\Psi)\left\{\cos^2\phi |A|^2 + \sin^2\phi |B|^2 - \sin2\phi \, {\rm Re}(A\cdot B^*)\right\},
\label{pdf0}
\\
& \ \ \ \ \ \ \ \ \ \ \ \  \ \ \ \ \  \  \ \ \ \ \ \ \ \ \ \ \ \  \ \ \ \ \  \  \ \ \ \ \ \ \ \ \ \ \ \  \ \ \ \ \  \  \ \ \ \ \ \ \ \ \ \ \ \  \ \ \ \ \  \  \ \ \ \ \ \ \ \ \ \ \ \  \ \ \ \ \  ({\rm general \ process})\nn
\end{align}
 where
\begin{align}
N_{{\rm nor}} &= \frac{2}{3} \int\int {\rm d}s_{12}{\rm d}s_{13}(|A|^2 + |B|^2).\nn
\end{align}
From Eq.~(\ref{pdfA}), we find that the $\cos\Psi$ asymmetry is proportional to both the helicity asymmetry $P_A$ and 
 the term ${\rm Im}(A\cdot B^*)/N_{\rm nor}$, the latter of which is non-zero only with the strong phase.
$\cos\Psi$ is a na\"ive T-odd quantity, and its expectation value is non-zero only with the strong phase, 
 in accordance with what is stated in Section~1.

Once the function ${\rm Im}(A\cdot B^*)/N_{\rm nor}$ is known,
 one can measure $P_A$ using asymmetry of the number of events with $\cos\Psi>0$ and $\cos\Psi<0$.
Conversely, if the helicity asymmetry of $a_1^-$ is known a priori,
 one can determine ${\rm Im}(A\cdot B^*)/N_{\rm nor}$ by measuring 
 asymmetry of the number of events with $\cos\Psi>0$ and $\cos\Psi<0$.
This is indeed feasible in $\tau_L^-\to\nu_\tau a_1^-(\to\pi^-\pi^-\pi^+)$ process, for which the $a_1^-$ helicity is theoretically calculable
 as done in Section~2.
To be specific, we write the differential decay rate
 of the $\tau_L^-\to\nu_\tau a_1^-(\to\pi^-\pi^-\pi^+)$ process
 in terms of $\frac{1}{\Gamma} \frac{{\rm d}\Gamma_{\lambda_{\rm lab}}}{{\rm d}z}$ Eq.~(\ref{normalizeddiffrate}) for $\tau_L^-$ as
\begin{align}
&\frac{1}{\Gamma\left(\tau_L^-\to \nu_\tau a_1^-(\to\pi^-\pi^-\pi^+)\right)}
\frac{{\rm d}^5\Gamma\left(\tau_L^-\to \nu_\tau a_1^-(\to\pi^-\pi^-\pi^+)\right)}{{\rm d}\cos\Psi {\rm d}\phi \,{\rm d}s_{13} {\rm d}s_{23} \,{\rm d}z}\nonumber\\
&=\left(\frac{1}{\Gamma}\frac{{\rm d}\Gamma_+}{{\rm d}z}+\frac{1}{\Gamma}\frac{{\rm d}\Gamma_-}{{\rm d}z}\right)
 \frac{1}{4\pi N_{{\rm nor}}}\left[|A|^2 + |B|^2 - (1-\cos^2\Psi) \left\{\cos^2\phi |A|^2 + \sin^2\phi |B|^2 - \sin2\phi \, {\rm Re}(A\cdot B^*)\right\}\right]
\label{taudecayT}\\
&+\left(\frac{1}{\Gamma}\frac{{\rm d}\Gamma_+}{{\rm d}z}-\frac{1}{\Gamma}\frac{{\rm d}\Gamma_-}{{\rm d}z}\right)
\frac{1}{4\pi N_{{\rm nor}}} 2\cos\Psi \ {\rm Im}(A\cdot B^*)
\label{taudecayA} \\
&+\frac{1}{\Gamma}\frac{{\rm d}\Gamma_0}{{\rm d}z}
\frac{1}{4\pi N_{{\rm nor}}}2(1-\cos^2\Psi)\left\{\cos^2\phi |A|^2 + \sin^2\phi |B|^2 - \sin2\phi \, {\rm Re}(A\cdot B^*)\right\}.
\label{taudecay0}
\end{align}
Since $\frac{1}{\Gamma} \frac{{\rm d}\Gamma_{\lambda_{\rm lab}}}{{\rm d}z}$ can be computed theoretically, 
 it is possible to determine ${\rm Im}(A\cdot B^*)/N_{\rm nor}$ from the $\cos\Psi$ asymmetry of $\tau_L^-\to\nu_\tau a_1^-(\to\pi^-\pi^-\pi^+)$ events.
A problem is that when we use $pp\to W^\mp \to \nu\,\tau^{\mp}(\to \nu\,\pi^\mp\pi^\mp\pi^\pm)$ events to collect $\tau_L^-$, it is difficult to reconstruct $z$, since two neutrinos contribute to the missing transverse momentum.
In this paper, we evade the reconstruction of $z$ by exploiting a positive correlation between $z$ and $a_1^-$'s transverse mass $M_T$.
We impose a tight selection cut on $M_T$ and thereby select events with large $z$.
${\rm Im}(A\cdot B^*)/N_{\rm nor}$ is determined from the $\cos\Psi$ asymmetry of events, divided by
 the convolution of theoretically calculated
 $\frac{1}{\Gamma} \frac{{\rm d}\Gamma_+}{{\rm d}z}-\frac{1}{\Gamma} \frac{{\rm d}\Gamma_-}{{\rm d}z}$
 and the reweighting function of $z$ under given selection cuts (the reweighting function is obtainable from a Monte Carlo simulation).

An advantage of the above method is that, for $\tau_L^-$ and for large $z$,
$\left|\frac{1}{\Gamma} \frac{{\rm d}\Gamma_+}{{\rm d}z}-\frac{1}{\Gamma} \frac{{\rm d}\Gamma_-}{{\rm d}z}\right|$
is maximized (see the left panel of Fig.~\ref{fracLR}).
Hence, the $\cos\Psi$ asymmetry is maximized
and the statistical uncertainty in the determination of ${\rm Im}(A\cdot B^*)/N_{\rm nor}$ is reduced.

We comment that, for $\tau_L^-$ and for $z\to1$, $\frac{1}{\Gamma} \frac{{\rm d}\Gamma_+}{{\rm d}z}-\frac{1}{\Gamma} \frac{{\rm d}\Gamma_-}{{\rm d}z}$
 quickly approaches to its value at $z=1$
 and is insensitive to the precise value of $z$.
Hence, once we collect large-$z$ events, the convolution of $\frac{1}{\Gamma} \frac{{\rm d}\Gamma_+}{{\rm d}z}-\frac{1}{\Gamma} \frac{{\rm d}\Gamma_-}{{\rm d}z}$ and the reweighting function of $z$
 is not affected by details of the reweighting function,
 which reduces the systematic uncertainty associated with the estimation of the reweighting function.
 However, confirming this reduction of the systematic uncertainty is beyond the scope of the present paper.
\\

\section{Method to determine ${\rm Im}(A\cdot B^*)/N_{\rm nor}$}

\subsection{Method} 

${\rm Im}(A\cdot B^*)/N_{\rm nor}$ satisfies the following relation stemming from Eq.~(\ref{taudecayA}):
\begin{align}
&\frac{{\rm Im}(A\cdot B^*)}{N_{\rm nor}}(Q^2,s_{13},s_{23})\ N(Q^2)
\left(\left.\frac{1}{\Gamma}\frac{{\rm d}\Gamma_+}{{\rm d}z}-\frac{1}{\Gamma}\frac{{\rm d}\Gamma_-}{{\rm d}z}\right\vert_{z,Q^2,\,{\rm for}\,\tau_L^-}\right)
\nn\\
&=\frac{{\rm d}^3N}{{\rm d}z{\rm d}s_{13}{\rm d}s_{23}}(\cos\Psi>0;\, z,Q^2,s_{13},s_{23})-
\frac{{\rm d}^3N}{{\rm d}z{\rm d}s_{13}{\rm d}s_{23}}(\cos\Psi<0;\, z,Q^2,s_{13},s_{23})
\label{exp}
\end{align}
 where ${\rm d}^3N(\cos\Psi>0;z,Q^2,s_{13},s_{23})/{\rm d}z{\rm d}s_{13}{\rm d}s_{23}$ denotes the number of
 $pp\to W^\mp \to \nu\,\tau^{\mp}(\to \nu\,\pi^\mp\pi^\mp\pi^\pm)$ events with $\cos\Psi>0$
 per $z,s_{13},s_{23}$ for fixed $Q^2$ (and likewise for $\cos\Psi<0$),
 and $N(Q^2)$ is the total number of events for any $z,s_{13},s_{23}$ for fixed $Q^2$ given by
\begin{align}
&N(Q^2)=
\nn\\
&\int{\rm d}z\int\int{\rm d}s_{12}{\rm d}s_{13}\left\{\frac{{\rm d}^3N}{{\rm d}z{\rm d}s_{13}{\rm d}s_{23}}(\cos\Psi>0;\, z,Q^2,s_{13},s_{23})+\frac{{\rm d}^3N}{{\rm d}z{\rm d}s_{13}{\rm d}s_{23}}(\cos\Psi<0;\, z,Q^2,s_{13},s_{23})\right\}.
\label{ntot}
\end{align}
Remind that $N(Q^2)$ and ${\rm Im}(A\cdot B^*)/N_{\rm nor}$ do not depend on $z$.

In real experiments, we cannot measure the right hand side of Eq.~(\ref{exp}), since we do not reconstruct $z$.
Instead, we propose to measure the following quantity:
\begin{align}
\frac{{\rm d}^2N_{\rm cut}}{{\rm d}s_{13}{\rm d}s_{23}}(\cos\Psi>0;\,Q^2,s_{13},s_{23})-\frac{{\rm d}^2N_{\rm cut}}{{\rm d}s_{13}{\rm d}s_{23}}(\cos\Psi<0;\,Q^2,s_{13},s_{23}),
\label{exp2}
\end{align}
 where ${\rm d}^2N_{\rm cut}(\cos\Psi>0;\,Q^2,s_{13},s_{23})/{\rm d}s_{13}{\rm d}s_{23}$ denotes the number of events with $\cos\Psi>0$
 per $s_{13},s_{23}$ for fixed $Q^2$, \textit{under given selection cuts} on 
the absolute values of the transverse momentum and pseudo-rapidity of $a_1$, 
the absolute value of the missing transverse momentum, and $a_1$'s transverse mass.
When one flips the sign of all three-momentum vectors, 
 the above selection cuts are invariant and so are $z,Q^2,s_{13},s_{23}$ and $\phi$.
On the other hand, $\cos\Psi$, which is proportional to the triple vector product of $\vec{p}_3,\ \vec{p}_1$ and the $a_1$ boost direction, flips its sign.
As a result, if $z,Q^2,s_{13},s_{23},\phi$ are the same, the above selection cuts do not discriminate an event with $\cos\Psi=c$ and one with $\cos\Psi=-c$.
Therefore, we can recast Eq.~(\ref{exp2}) in the form,
\begin{align}
&(\ref{exp2})=\int{\rm d}z \ f_{\rm cut}(z,Q^2,s_{13},s_{23})
\nn\\
&\times\left\{\frac{{\rm d}^3N}{{\rm d}z{\rm d}s_{13}{\rm d}s_{23}}(\cos\Psi>0;\, z,Q^2,s_{13},s_{23})-
\frac{{\rm d}^3N}{{\rm d}z{\rm d}s_{13}{\rm d}s_{23}}(\cos\Psi<0;\, z,Q^2,s_{13},s_{23})\right\}
\label{exp2-2}
\end{align}
where $f_{\rm cut}(z,Q^2,s_{13},s_{23})$ denotes the fraction of events with specific values of $z,Q^2,s_{13},s_{23}$
that pass the selection cuts, which is common for $\cos\Psi>0$ and $\cos\Psi<0$.
From Eqs.~(\ref{exp}),(\ref{exp2-2}) and $z$-independence of $N(Q^2)$ and ${\rm Im}(A\cdot B^*)/N_{\rm nor}$,
 we have
\begin{align}
&\frac{{\rm Im}(A\cdot B^*)}{N_{\rm nor}}(Q^2,s_{13},s_{23}) \ N(Q^2)\int{\rm d}z \ f_{\rm cut}(z,Q^2,s_{13},s_{23})
\left(\left.\frac{1}{\Gamma}\frac{{\rm d}\Gamma_+}{{\rm d}z}-\frac{1}{\Gamma}\frac{{\rm d}\Gamma_-}{{\rm d}z}\right\vert_{z,Q^2,\,{\rm for}\,\tau_L^-}\right)
\nn\\
&=\frac{{\rm d}^2N_{\rm cut}}{{\rm d}s_{13}{\rm d}s_{23}}(\cos\Psi>0;\,Q^2,s_{13},s_{23})-\frac{{\rm d}^2N_{\rm cut}}{{\rm d}s_{13}{\rm d}s_{23}}(\cos\Psi<0;\,Q^2,s_{13},s_{23}).
\label{exp3}
\end{align}
The right hand side is measured in experiments.
As for the left hand side, $N(Q^2)$ is known from the $pp\to W^\mp$ total cross section and branching fractions
 of $W^-\to\tau^-\bar{\nu}_\tau$ and $\tau^-\to\pi^-\pi^-\pi^+\nu_\tau$ decays.
$\frac{1}{\Gamma} \frac{{\rm d}\Gamma_{\lambda_{\rm lab}}}{{\rm d}z}$ has been calculated theoretically in Section~2.
$f_{\rm cut}(z,Q^2,s_{13},s_{23})$ can be evaluated with a Monte Carlo simulation by exploiting generator-level information on $z$.
Therefore, it is possible to determine ${\rm Im}(A\cdot B^*)/N_{\rm nor}$.
\\

For practical purposes, it is convenient to use $w(z,Q^2,s_{13},s_{23})$ defined below, in place of $f_{\rm cut}(z,Q^2,s_{13},s_{23})$:
\begin{align}
w(z,Q^2,s_{13},s_{23})&=\frac{N(Q^2)\,f_{\rm cut}(z,Q^2,s_{13},s_{23})}{N_{\rm cut}(Q^2)}
\label{wdef}
\end{align}
 where $N_{\rm cut}(Q^2)$ is the total number of events for fixed $Q^2$ that pass the selection cuts,
\begin{align}
&N_{\rm cut}(Q^2)=
\int\int{\rm d}s_{12}{\rm d}s_{13}\left\{\frac{{\rm d}^2N_{\rm cut}}{{\rm d}s_{13}{\rm d}s_{23}}(\cos\Psi>0;\,Q^2,s_{13},s_{23})+\frac{{\rm d}^2N_{\rm cut}}{{\rm d}s_{13}{\rm d}s_{23}}(\cos\Psi<0;\,Q^2,s_{13},s_{23})\right\}.
\label{ncuttot}
\end{align}
$w(z,Q^2,s_{13},s_{23})$ is interpreted as the reweighting of events with specific values of $z,s_{13},s_{23},Q^2$ due to the selection cuts,
 which is again common for $\cos\Psi>0$ and $\cos\Psi<0$.
In terms of $w(z,Q^2,s_{13},s_{23})$, Eq.~(\ref{exp3}) is recast in the form,
\begin{align}
&\frac{{\rm Im}(A\cdot B^*)}{N_{\rm nor}}(Q^2,s_{13},s_{23}) \ N_{\rm cut}(Q^2)\int{\rm d}z \ w(z,Q^2,s_{13},s_{23})
\left(\left.\frac{1}{\Gamma}\frac{{\rm d}\Gamma_+}{{\rm d}z}-\frac{1}{\Gamma}\frac{{\rm d}\Gamma_-}{{\rm d}z}\right\vert_{z,Q^2,\,{\rm for}\,\tau_L^-}\right)
\nn\\
&=\frac{{\rm d}^2N_{\rm cut}}{{\rm d}s_{13}{\rm d}s_{23}}(\cos\Psi>0;\,Q^2,s_{13},s_{23})-\frac{{\rm d}^2N_{\rm cut}}{{\rm d}s_{13}{\rm d}s_{23}}(\cos\Psi<0;\,Q^2,s_{13},s_{23}).
\label{exp4}
\end{align}
\\

In Section~4.2, we generate detector-level Monte Carlo events for the 14~TeV LHC and impose selection cuts on them.
Using these events, in Section~4.3, we evaluate $w(z,Q^2,s_{13},s_{23})$.
In Section~4.4, we estimate statistical uncertainty in a measurement of the right hand side of Eq.~(\ref{exp4}) with 300~fb$^{-1}$ of data.
\\

\subsection{Monte Carlo Event Generation and Selection Cuts}

Using MadGraph5\underline{\ }aMC@NLO~\cite{mg} with the TauDecay package~\cite{taudecay}, 
we generate parton-level events for the process (charged-conjugated process is also considered),
\bea
pp \to W^-(\to\bar{\nu}_\tau\tau^- (\to\nu_\tau \pi^- \pi^- \pi^+)) \ + \ 0,1,2 \ {\rm parton(s)}
\eea
 for $\sqrt{s}=14$~TeV $pp$ collisions.
Then, we use PYTHIA8~\cite{pythia} to simulate parton showering.
The groups of events with 0, 1, 2 parton(s) are matched with MLM-matching~\cite{mlm} algorithm.

For the events generated, we use the Delphes3 program~\cite{delphes} 
to simulate the CMS detector effects, considering $|\eta_{\tau{\rm \mathchar`-jet}}| < 2.5$, $p_{\tau{\rm \mathchar`-jet}T} > 1$~GeV and 60\% tagging efficiency for the identification of a three-prong $\tau$-jet.
We reconstruct an $a_1^-$ meson from a three-prong $\tau$-jet
 by requiring that the three charged tracks have charges summed to that of $a_1^-$ and that
 the invariant mass (calculated by assuming that each charged track is a pion) be less than 2 GeV.
The variables $\cos\Psi$, $s_{13}$, $s_{23}$ are calculated as described in Section 3. 
\\

We impose the following selection cuts on the above samples:

\begin{itemize}

\item Event must contain exactly one reconstructed $a_1^-$ meson.

\item Missing transverse momentum must satisfy $\mpt_T>25$~GeV.

\item Invariant mass of the $a_1^-$ must satisfy 1.26~GeV$>\sqrt{Q^2}>$1.20~GeV.

\end{itemize}

Additionally, we impose

\begin{itemize}

\item Transverse mass for the $a_1^-$,
\begin{align} 
M_T&=\sqrt{2\vert\vec{\mpt}_T\vert\sqrt{\vert\vec{p}_{a_1T}\vert^2+Q^2} (1-\cos \phi_{a_1\mpt_T})}
\end{align}
 where $\vec{p}_{a_1T}$ denotes the transverse momentum of the $a_1^-$, and $\phi_{a_1\mpt_T}$ is the azimuthal angle between the $a_1^-$ and the missing transverse momentum,
 must satisfy either $M_T>50$~GeV, 60~GeV or 70~GeV.

\end{itemize}

The number of the sum of $W^+$ and $W^-$ events with 300~fb$^{-1}$ of data at each stage of event selection and for each $M_T$ cut is tabulated in Table~\ref{number}.
\begin{table}[H]
\begin{center}
  \caption{Number of the sum of $pp\to W^\mp \to \nu\,\tau^{\mp}(\to \nu\,\pi^\mp\pi^\mp\pi^\pm)$ events 
  at the 14~TeV LHC with 300~fb$^{-1}$ of data, after each selection cut.}
\begin{tabular}[H]
{|c|c|} \hline
Selection cut & Number of $W^+$ and $W^-$ events \\ \hline\hline
One reconstructed $a_1^-$ and $\mpt_T>25$~GeV & 40.7$\times10^6$ \\
\& 1.26~GeV$>\sqrt{Q^2}>$1.20~GeV & $5.12\times10^6$ \\ 
\& $M_T>50$~GeV & $4.47\times10^6$ \\ 
\& $M_T>60$~GeV & $3.00\times10^6$ \\ 
\& $M_T>70$~GeV & $1.43\times10^6$ \\ \hline
\end{tabular}
  \label{number}
  \end{center}
\end{table}

\subsection{Estimation of $w(z,Q^2,s_{13},s_{23})$}

\begin{figure}[H]
\begin{center}
\includegraphics[width=11cm]{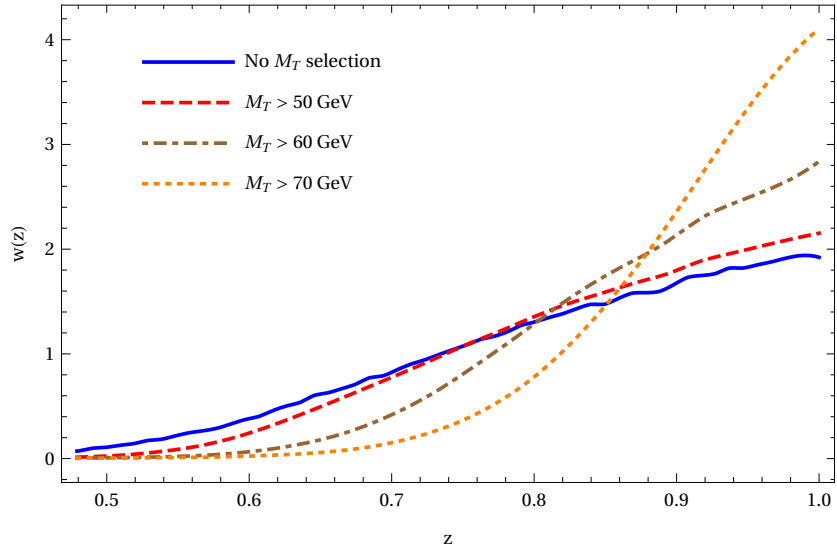}
\caption{$w(z)$, the reweighting of events with specific value of $z$ due to the selection cuts of Section~4.2
(which does not depend on $s_{13},s_{23}$ and is uniform in the bin of $Q^2$).
The cases with $M_T>50$~GeV, 60~GeV and 70~GeV and the case without $M_T$ cut are plotted.
}
\label{wplots}
\end{center}
\end{figure}

Using the above event samples,
we estimate $w(z,Q^2,s_{13},s_{23})$ Eq.~(\ref{wdef}), which is the reweighting of events with specific values of $z,s_{13},s_{23},Q^2$ due to the selection cuts, which is common for $\cos\Psi>0$ and $\cos\Psi<0$.
In fact, the selection cuts of Section~4.2 do not distort the distributions of $Q^2,s_{13},s_{23}$ and we simply have
\begin{align}
w(z,Q^2,s_{13},s_{23})&=w(z) \ \ \ \ \ \ {\rm for} \ 1.26 \ {\rm GeV}>\sqrt{Q^2}>1.20\ {\rm GeV}
\nn\\
w(z,Q^2,s_{13},s_{23})&=0  \ \ \ \ \ \ \ \ \ \ {\rm otherwise}\nn
\end{align}
In Fig.~\ref{wplots}, we present $w(z)$ for the selection cuts with $M_T>50$~GeV, 60~GeV and 70~GeV and for the case without $M_T$ cut.
We have confirmed that large-$z$ events are efficiently collected with a tight $M_T$ cut such as $M_T>70$~GeV.
\\

\subsection{Statistical Uncertainty}
 
We estimate statistical uncertainty in a measurement of the right hand side of Eq.~(\ref{exp4}) with 300~fb$^{-1}$ of data.

\begin{figure}[H]
\begin{center}
\includegraphics[width=8cm]{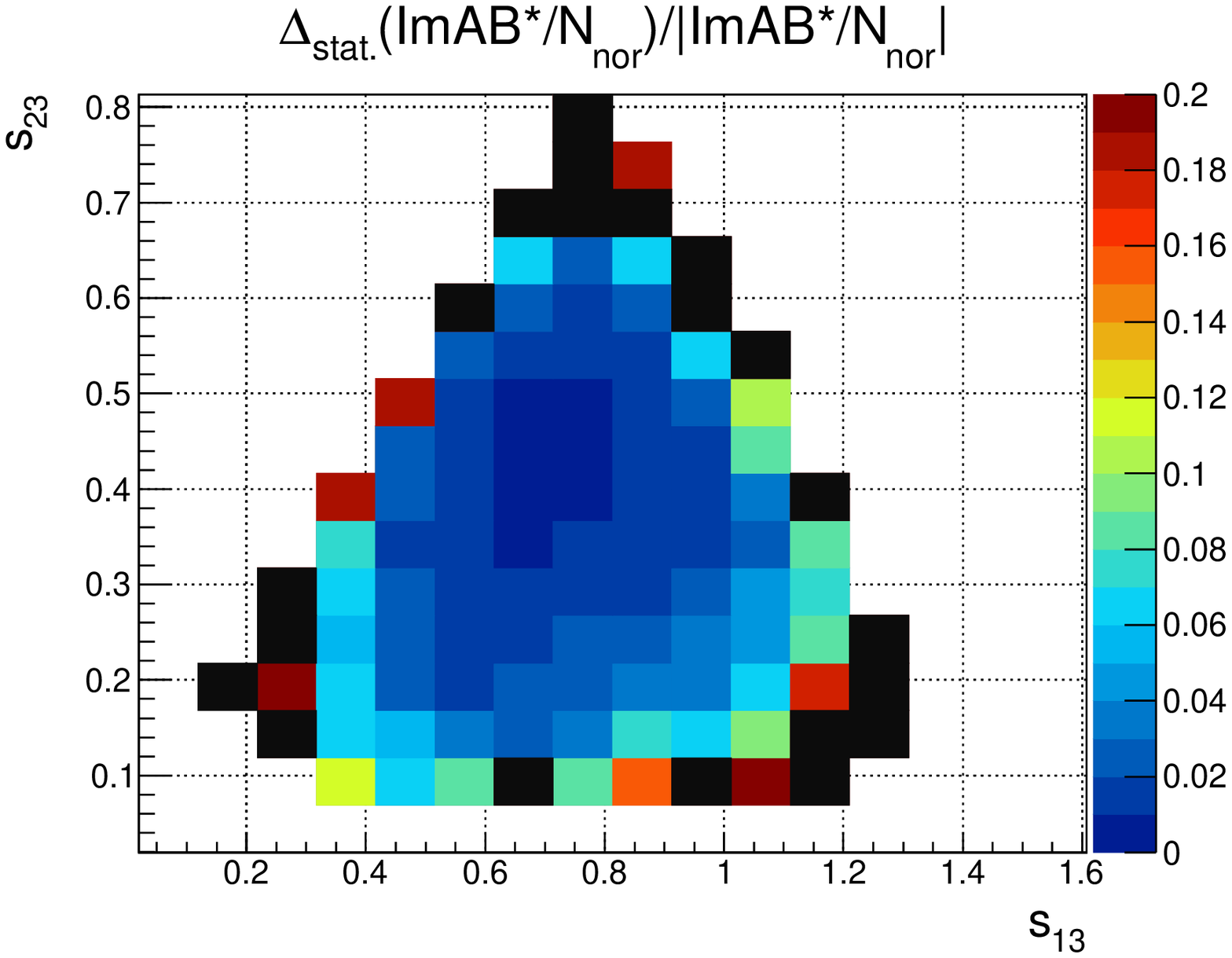}
\includegraphics[width=8cm]{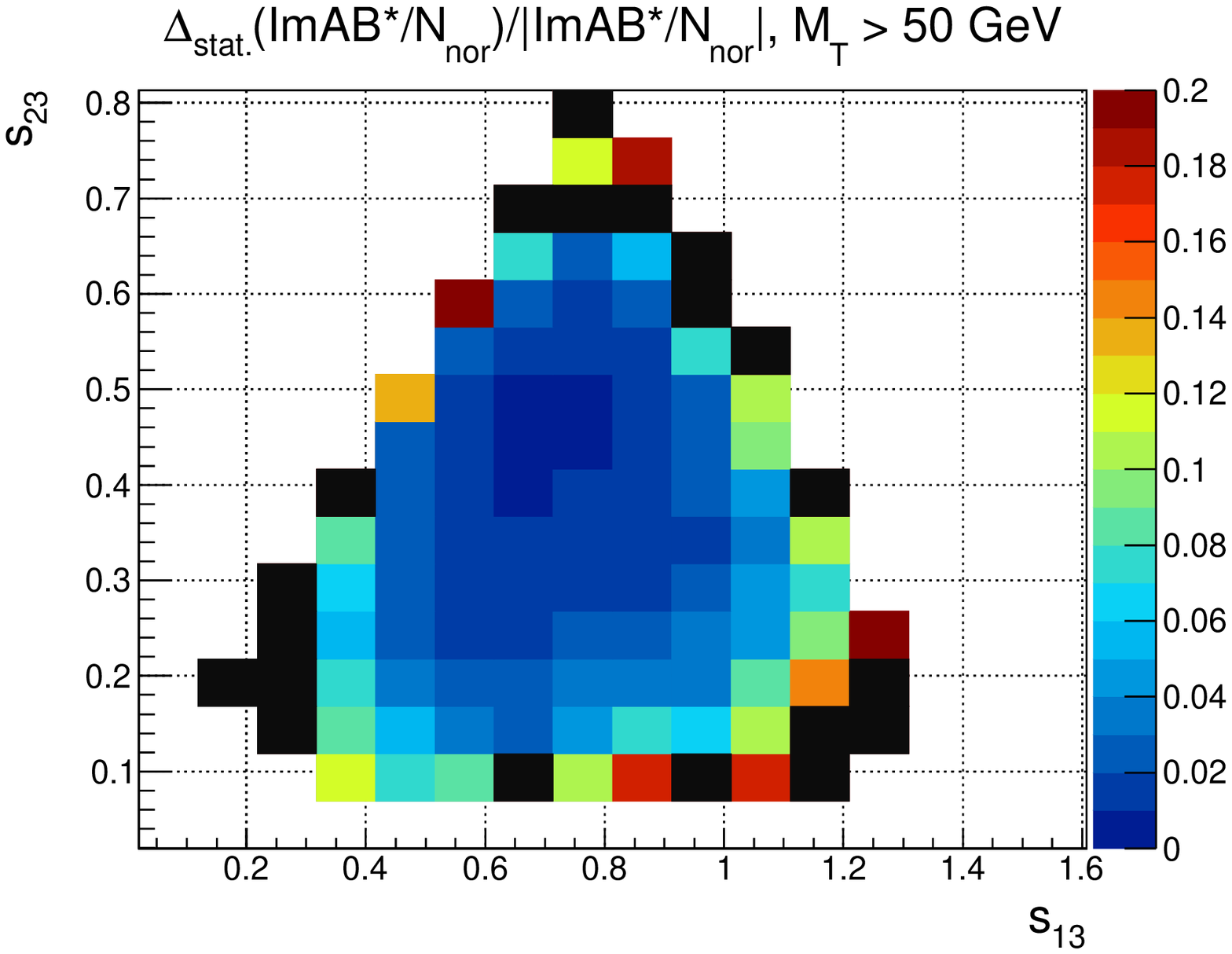}
\\
\includegraphics[width=8cm]{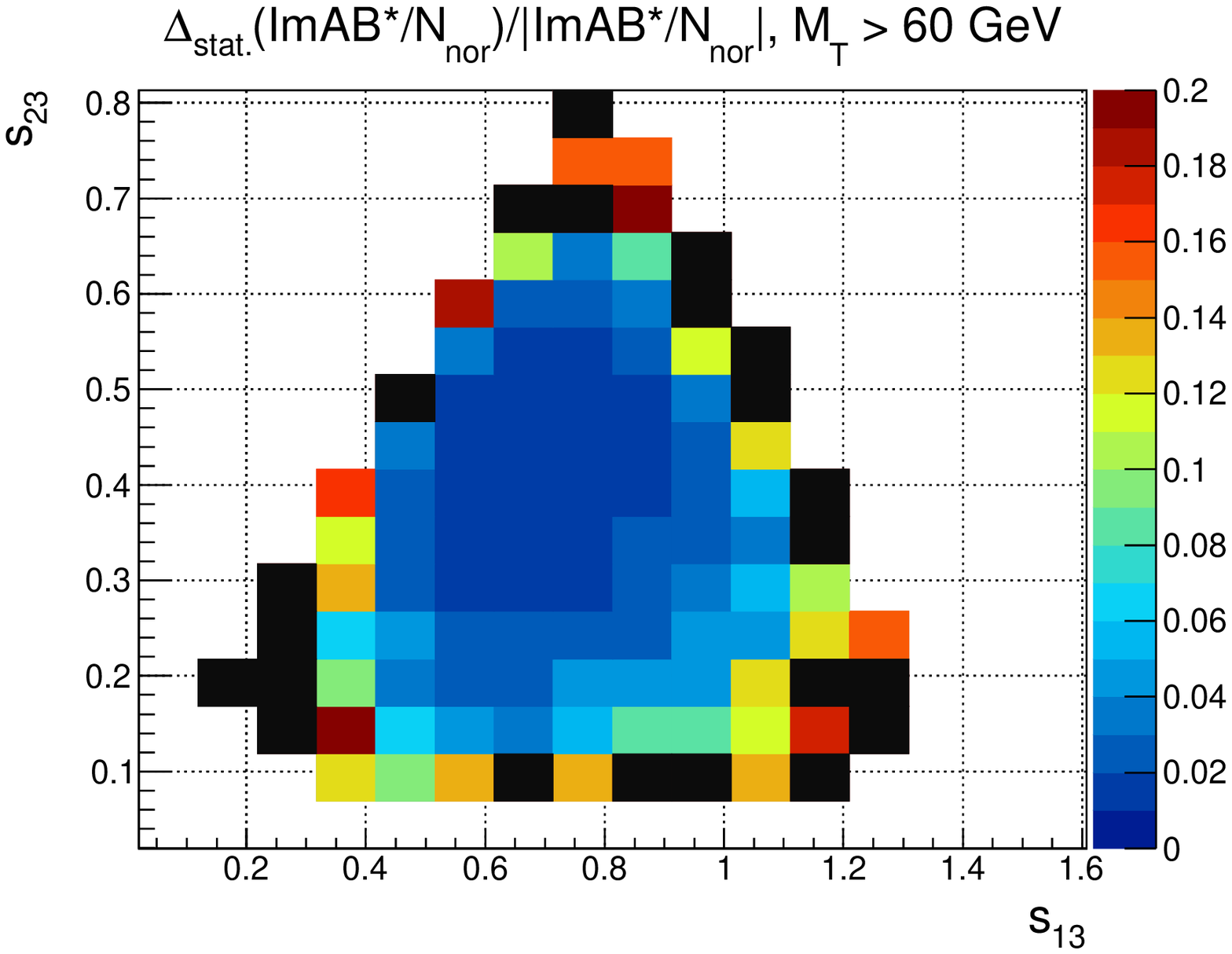}
\includegraphics[width=8cm]{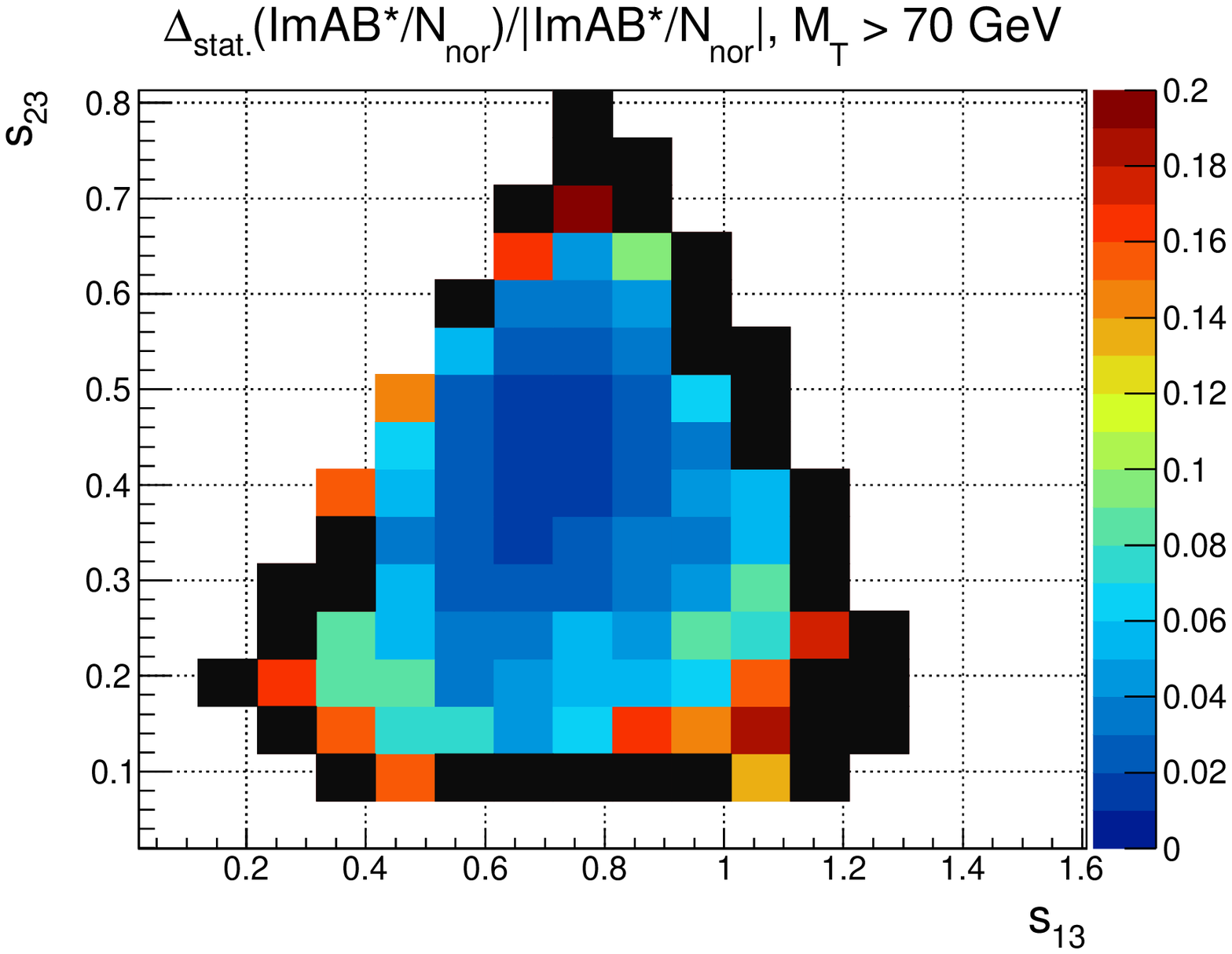}
\caption{Relative statistical uncertainty in the determination of ${\rm Im}(A\cdot B^*)/N_{\rm nor}$
at the 14~TeV LHC with 300~fb$^{-1}$ of data, in each bin of $(s_{13},s_{23})$ (in units of GeV$^2$)
 with 1.26~GeV$>\sqrt{Q^2}>$1.20~GeV.
The upper-left, upper-right, lower-left and lower-right panels correspond to the event selection without $M_T$ cut, with $M_T>50$~GeV,
with $M_T>60$~GeV, and with $M_T>70$~GeV, respectively.
}
\label{statplots}
\end{center}
\end{figure}

The statistical uncertainty is given as follows:
Let $\delta N_+$ ($\delta N_-$) denote the number of events after the selection cuts of Section~4.2,
in a bin of $Q^2,s_{13},s_{23}$ with $\cos\Psi>0$ $(\cos\Psi<0)$.
If the bins are sufficiently narrow, the right hand side of Eq.~(\ref{exp4}) is approximated by
\bea
\delta N_+-\delta N_-,
\eea
 for which the ratio of the statistical uncertainty over its value is given by
\bea
\frac{\Delta_{\rm stat}(\delta N_+-\delta N_-)}{\delta N_+-\delta N_-}=\frac{\sqrt{\delta N_++\delta N_-}}{\delta N_+-\delta N_-}.
\eea
This corresponds the relative statistical uncertainty in the determination of ${\rm Im}(A\cdot B^*)/N_{\rm nor}$, and hence we obtain
\bea
\frac{\Delta_{\rm stat}({\rm Im}(A\cdot B^*)/N_{\rm nor})}{{\rm Im}(A\cdot B^*)/N_{\rm nor}}=\frac{\sqrt{\delta N_++\delta N_-}}{\delta N_+-\delta N_-}.
\label{stat}
\eea
In Fig.~\ref{statplots}, we present the relative statistical uncertainty Eq.~(\ref{stat}) in each bin of $(s_{13},s_{23})$
 with 1.26~GeV$>\sqrt{Q^2}>$1.20~GeV,
 for various $M_T$ cuts, at the 14~TeV LHC with 300~fb$^{-1}$ of data.
We observe that the relative statistical uncertainty is below 2\% in multiple bins of $(s_{13},s_{23})$,
 and so the determination of ${\rm Im}(A\cdot B^*)/N_{\rm nor})$ is feasible at the 14~TeV LHC with 300~fb$^{-1}$ of data,
 at least in light of statistics.

A tighter $M_T$ cut diminishes overall statistics, but it enhances the relative $\cos\Psi$ asymmetry $\left|\frac{\delta N_+-\delta N_-}{\delta N_++\delta N_-}\right|$,
because this asymmetry is proportional to $\left|\frac{1}{\Gamma} \frac{{\rm d}\Gamma_+}{{\rm d}z}-\frac{1}{\Gamma} \frac{{\rm d}\Gamma_-}{{\rm d}z}\right|$ and the latter is largest for $z\sim1$.
Nevertheless, we do not find improvement in relative statistical uncertainty with tigher $M_T$ cuts.
This is because in the present simulation, the loss of overall statistics is more significant than the enhancement of the relative $\cos\Psi$ asymmetry.

\section{Summary}

We have presented a method to extract information about the strong phase of the $a_1^- \to \pi^-\pi^-\pi^+$ decay amplitude
 necessary for the $a_1$ helicity measurement.
Our method utilizes $W \to \nu \tau(\to \nu \pi^\mp \pi^\mp \pi^\pm)$ events, for which the $a_1$ helicity is theoretically calculable.
The method has an advantage that $a_1^-$ mesons from $\tau_L^-$ decays with large boost (i.e. with $z=E_{a_1}/E_\tau \sim1$ in the laboratory frame)
have nearly maximal helicity asymmetry and thus most reflect the strong phase.
We have revisited the theoretical calculation of the $a_1^-$ helicity in the laboratory frame 
 in $W^- \to \bar{\nu}_\tau \tau^-(\to \nu_\tau a_1^-)$ process.
We have formulated the differential decay rate of polarized $a_1$ mesons, where the information about the strong phase necessary for the helicity measurement is encapsulated by the term ${\rm Im}(A\cdot B^*)/N_{\rm nor}$.
Finally, we have proposed a method to determine ${\rm Im}(A\cdot B^*)/N_{\rm nor}$ from $pp\to W \to \nu \tau(\to \nu \pi^\mp \pi^\mp \pi^\pm)$ events, and by estimating the statistical uncertainty at the 14~TeV LHC with 300~fb$^{-1}$ of data, we have revealed that
this method is feasible at least in light of statistics.
\\

\section*{Acknowledgement}

KH thanks Bernd Kniehl and members of II. Institut f\"ur Theoretische Physik, Universit\"at Hamburg, 
where part of the work was done. 
KH and TY thank Korea Institute for Advanced Study, where part of the work was done.
DY thanks the KEK Theory Center for the warm hospitality, and Qiang Li for continuous supports. The numerical simulations were performed on the Computing Cluster at the Center of High Energy Physics, Tsinghua University. 
This work is partially supported by Scientific Grants by the Ministry
of Education, Culture, Sports, Science and Technology of Japan, Nos.~19K147101 (TY) and 18H03708 (HI).
\\

\section*{Appendix: Wigner rotation}

The rotation between two angular momentum quantization axes of a massive particle in its rest frame, 
where the two axes are chosen along its three momentum in different Lorentz frames, 
is called Wigner rotation, following his historical paper Ref.~\cite{wigner}. 
In this appendix, we give its simple derivation because it is not widely known to contemporary high energy physicists. 

\begin{figure}[H]
\begin{center}
\includegraphics[width=8cm]{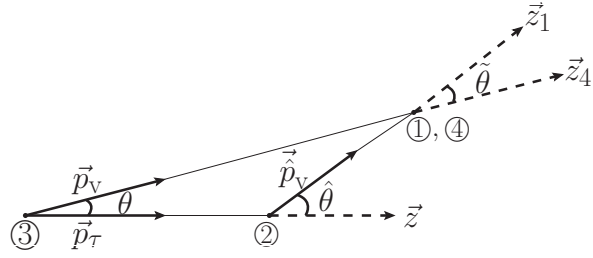}
\caption{Wigner rotation angle $\tilde{\theta}$. 
\textcircled{\raisebox{-0.4mm}{1}} and \textcircled{\raisebox{-0.4mm}{4}} 
are vector meson rest frames, 
where $\vec{z}_1$ is chosen along the meson momentum direction 
in the $\tau$ rest frame \textcircled{\raisebox{-0.4mm}{2}}, 
whereas $\vec{z}_4$ is along the meson momentum direction 
in the laboratory frame \textcircled{\raisebox{-0.4mm}{3}}. }
\label{Fig:Wigrot}
\end{center}
\end{figure}
In Fig.~\ref{Fig:Wigrot}, we show all the relevant Lorentz frames: 
The Lorentz frame \textcircled{\raisebox{-0.4mm}{1}} and \textcircled{\raisebox{-0.4mm}{4}} 
is a vector boson rest frame, 
where angular-momentum-quantization axis $\vec{z}_1$ is chosen along the vector boson's three-momentum 
in the $\tau$ rest frame \textcircled{\raisebox{-0.4mm}{2}},
while axis $\vec{z}_4$ is chosen along the vector boson's three-momentum 
in the laboratory frame \textcircled{\raisebox{-0.4mm}{3}}. 
The $\tau$ rest frame \textcircled{\raisebox{-0.4mm}{2}} is obtained from 
the laboratory frame \textcircled{\raisebox{-0.4mm}{3}} 
by a boost along the $\tau$ momentum direction $\vec{z}$ 
in the laboratory frame. 
Since the above successive transformations 
are all on the $(z, x)$ plane,
 the quantization axis $\vec{z}_1$ in the vector boson rest frame is recovered 
 by a rotation by $\tilde{\theta}$ about the common $y$-axis, namely, we have
\begin{align}
1 = R_y (\tilde{\theta}) B_{z_4}^{-1} (y_{\rm v}) R_y (\theta) B_z (y_\tau) R_y (-\hat{\theta}) B_{z_1} (\hat{y}_{\rm v})\,,\tag{A.1}\label{Eq:A1}
\end{align}
where $R_{y} (\theta) = e^{-i J_2 \theta}$ denotes rotation about the common $y$-axis 
and $B_z (y) = e^{-i K_3 y}$ denotes boost along the corresponding $z$-direction depicted in Fig.~\ref{Fig:Wigrot}. 
Only two generators of the Lorentz transformations appear in Eq.~\eqref{Eq:A1}, 
whose non-zero components are 
$(J_2)_{jk} = -i \epsilon_{2 jk}$, $(K_3)_{0k} = (K_3)_{k0} = i \delta_{k3}$. 
In Eq.~(\ref{Eq:A1}), $\hat{\theta}$ denotes the angle between $z$-axis and the vector boson three-momentum in the $\tau$ rest frame,
 $\theta$ denotes the angle between $z$-axis and the vector boson's three-momentum in the laboratory frame,
 and $\tilde{\theta}$ is the Wigner rotation angle we want to derive.
The rapidity along each direction satisfies
\begin{align}
\tanh \hat{y}_{\rm v} &= \hat{p}_{\rm v}/m_{\rm v} = \hat{k}\,,\tag{A.2a}\label{Eq:A2a}\\
\tanh y_{\tau}             &= p_{\tau}/m_{\tau} = \gamma \beta\,,\tag{A.2b}\label{Eq:A2b}\\
\tanh y_{\rm v}           &= p_{\rm v}/m_{\rm v} 
= \sqrt{\gamma^2 (\beta \hat{\omega} + \hat{k} \cos \hat{\theta})^2 + (\hat{k} \sin \hat{\theta})^2}\,,\tag{A.2c}\label{Eq:A2c}
\end{align}
where $\hat{p}_{\rm v}$ denotes the vector boson's three-momentum in the $\tau$ rest frame,
　$p_{\rm v}$ denotes the vector boson's three-momentum in the laboratory frame,
and $p_{\tau}$ denotes the $\tau$ lepton's three-momentum in the laboratory frame.
Here, we have parametrized the $\tau$ lepton's four-momentum in the laboratory frame \textcircled{\raisebox{-0.4mm}{3}} as
\begin{align}
p^{\mu}_\tau = (E_\tau, 0, 0, p_\tau) = m_{\tau} (\gamma, 0, 0, \gamma \beta)\,.\tag{A.4}\label{Eq:A4}
\end{align}

$\hat{p}_{\rm v}$ and the corresponding energy are easily derived as
\begin{align}
\hat{E}_{\rm v} 
&= 
\frac{m_\tau}{2} \left( 1+a^2\right)
= 
m_{\rm v} \frac{1+a^2}{2a} 
= 
m_{\rm v} \hat{\omega} \,,\tag{A.3a}\label{Eq:A3a}\\
\hat{p}_{\rm v} 
&= 
\frac{m_\tau}{2} \left( 1-a^2\right)
= 
m_{\rm v} \frac{1-a^2}{2a}
= 
m_{\rm v} \hat{k} \,,\tag{A.3b}\label{Eq:A3b}
\end{align}
with $a = m_{\rm v}/m_{\tau}$.

The vector boson's four-momentum in the laboratory frame can be expressed as
\begin{align}
  E_{\rm v}   
  &= 
  m_{\rm v} \gamma \left( \hat{\omega} + \beta \hat{k} \cos \hat{\theta} \right)\,,\tag{A.5a}\label{Eq:A5a}\\
  p_{\rm v}^1 
  &= 
  m_{\rm v} \hat{k} \sin \hat{\theta}\,,\tag{A.5b}\label{Eq:A5b}\\
  p_{\rm v}^2 
  &= 
  0\,,\tag{A.5c}\label{Eq:A5c}\\
  p_{\rm v}^3 
  &= 
  m_{\rm v} \gamma \left( \hat{k} \cos \hat{\theta} + \beta \hat{\omega} \right)\,,\tag{A.5d}\label{Eq:A5d}
\end{align}
and hence $p_{\rm v}$ and $\tan \theta$ are derived as
\begin{align}
  p_{\rm v}
  =
  m_{\rm v} \sqrt{\gamma^2 \left(\hat{k} \cos \hat{\theta} + \beta \hat{\omega} \right)^2
           + \left(\hat{k} \sin \hat{\theta} \right)^2}\,,\tag{A.6}\label{Eq:A6}
\end{align}
and
\begin{align}
  \tan \theta
  =
  \frac{\hat{k} \sin \hat{\theta}}{\gamma \left(\hat{k} \cos \hat{\theta} + \beta \hat{\omega} \right)}\,.\tag{A.7}\label{Eq:A7}
\end{align}
Eq.~(\ref{Eq:A6})
determines the boost factor (\ref{Eq:A2c}).

Straightforward calculation gives
\begin{align}
R_y ( \theta )B_z ( y_{\tau} ) R_y( - \hat{\theta} ) B_{z_1} (\hat{y})
 =
\begin{pmatrix}
\gamma \left( \hat{\omega} + \hat{c} \hat{k} \beta \right) &\gamma \hat{s} \beta &0 &\gamma \left( \hat{k} + \hat{c} \hat{\omega} \beta \right)\\
\gamma s \left( \hat{c} \hat{k} + \hat{\omega} \beta \right) - c  \hat{s} \hat{k} &c \hat{c} + \gamma s \hat{s} &0 &\gamma s \left( \hat{c} \hat{\omega} + \hat{k} \beta \right) - c  \hat{s} \hat{\omega}\\
0 &0 &1 &0\\
\gamma c \left( \hat{c} \hat{k} + \hat{\omega} \beta \right) + s  \hat{s} \hat{k} &- s \hat{c} + \gamma c \hat{s} &0  &\gamma c \left( \hat{c} \hat{\omega} + \hat{k} \beta \right) + s  \hat{s} \hat{\omega}
\end{pmatrix}\,,\tag{A.8}\label{Eq:A8}
\end{align}
where $c (\hat{c})$ and $s (\hat{s})$ denote $\cos \theta (\cos \hat{\theta})$ and $\sin \theta (\sin \hat{\theta})$, respectively. 
From the definition of the Wigner rotation Eq.~(\ref{Eq:A1}), we get that Eq.~(\ref{Eq:A8}) should be equal to the following quantity:
\begin{align}
B_{z_4} (y_{\rm v}) R_y (-\tilde{\theta}) 
= 
\begin{pmatrix}
\cosh y_{\rm v} &\sinh y_{\rm v} \sin \tilde{\theta} &0 &\sinh y_{\rm v} \cos \tilde{\theta}\\
0 &\cos \tilde{\theta} &0 &-\sin \tilde{\theta}\\
0 &0 &1 &0\\
\sinh y_{\rm v} &\cosh y_{\rm v} \sin \tilde{\theta} &0 &\cosh y_{\rm v} \cos \tilde{\theta}
\end{pmatrix}\,.\tag{A.9}\label{Eq:A9}
\end{align}
Comparison of (1,1) components of (\ref{Eq:A8}) and (\ref{Eq:A9}) gives
\begin{align}
\cos \tilde{\theta} 
= 
\cos \theta \cos \hat{\theta} + \gamma \sin \theta \sin \hat{\theta}\,.\tag{A.10}\label{Eq:A10}
\end{align}
In the Gallilean transformation limit with $\gamma=1$, the Wigner rotation angle becomes
\begin{align}
\tilde{\theta} 
= 
\hat{\theta} - \theta\,,\tag{A.11}\label{Eq:A11}
\end{align}
as depicted in Fig.~\ref{Fig:Wigrot}.
In generic Lorentz transformations, 
the relation (\ref{Eq:A11}) no longer holds, 
and the Wigner rotation angle $\tilde{\theta}$ is obtained from Eq.(\ref{Eq:A10}).
By inserting (\ref{Eq:A7}) into (\ref{Eq:A10}), we obtain 
\begin{align}
\cos \tilde{\theta} = 
\frac{\beta (1+a^2) \cos \hat{\theta} + 1-a^2}{\sqrt{(\beta (1+a^2) + (1-a^2) \cos \hat{\theta})^2 + 
((1-a^2) \sin \hat{\theta}/\gamma)^2}}\,,\tag{A.12}\label{Eq:A12}
\end{align}
which gives Eq.~(\ref{wignerangle}) by noting $1/\gamma^2 = 1 - \beta^2$.

Let us give a few remarks on the Wigner rotation angle $\tilde{\theta}$. 
As is clear from the expression Eq.~\eqref{Eq:A6}, 
the Gallilean limit of $\tilde{\theta} = \hat{\theta} - \theta$ 
is recovered for $\gamma \to 1$. 
Since $\gamma > 1$, the relativistic correction gives $\tilde{\theta} < \hat{\theta} - \theta$. 
In the ultra-relativistic limit with $\gamma \to \infty~ (\beta \to 1)$, we find 
\begin{align}
\cos \tilde{\theta} 
\xrightarrow[\beta \to 1]{}
\frac{(1+a^2) \cos \hat{\theta} + 1-a^2}{(1-a^2) \cos \hat{\theta} + 1+a^2}\,,\tag{A.13}\label{Eq:A13}
\end{align}
which is a good approximation for a vector meson in the decay of a $\tau$ lepton coming from the decay of $W,Z$.
Finally, it is worth noting that the expression Eq.~\eqref{Eq:A9} gives the helicity conservation 
\begin{align}
\cos \tilde{\theta} 
\xrightarrow[a \to 0]{} 
1\,,\tag{A.14}\label{Eq:A14}
\end{align}
in the massless limit of the vector meson with $a = m_{\rm v}/m_{\tau} \to 0$. 
There is no rotation $(\tilde{\theta} = 0)$ in the massless limit, because the helicity of a massless particle 
is an invariant of Lorentz transformations.




\begin{thebibliography}{99}
 \bibitem{lhcb}
   R.~Aaij {\it et al.} [LHCb Collaboration],
  ``Observation of Photon Polarization in the $b\to s\gamma$ Transition,''
  Phys.\ Rev.\ Lett.\  {\bf 112}, no. 16, 161801 (2014)
  [arXiv:1402.6852 [hep-ex]].
  
   \bibitem{rujula}
    A.~De R\`ujula, R.~Petronzio and B.~E.~Lautrup,
  ``On the Challenge of Measuring the Color Charge of Gluons,''
  Nucl.\ Phys.\ B {\bf 146}, 50 (1978).
  
\bibitem{Yang:2004as} 
  H.~Yang {\it et al.} [Belle Collaboration],
  ``Observation of $B^+ \to K_1(1270) + \gamma$,''
  Phys.\ Rev.\ Lett.\  {\bf 94}, 111802 (2005)
  [hep-ex/0412039].
 
 \bibitem{gronau1}
    M.~Gronau, Y.~Grossman, D.~Pirjol and A.~Ryd,
  ``Measuring the photon polarization in $B \to K \pi \pi \gamma$,''
  Phys.\ Rev.\ Lett.\  {\bf 88}, 051802 (2002)
  [hep-ph/0107254].
 
 \bibitem{gronau2}
   M.~Gronau and D.~Pirjol,
  ``Photon polarization in radiative B decays,''
  Phys.\ Rev.\ D {\bf 66}, 054008 (2002)
  [hep-ph/0205065].
  
 \bibitem{kou1}
  E.~Kou, A.~Le Yaouanc and A.~Tayduganov,
  ``Determining the photon polarization of the $b \to s \gamma$ using the $B \to K_1(1270) \gamma \to (K \pi \pi) \gamma$ decay,''
  Phys.\ Rev.\ D {\bf 83}, 094007 (2011)
  [arXiv:1011.6593 [hep-ph]].
 
 \bibitem{kou2}
  A.~Tayduganov, E.~Kou and A.~Le Yaouanc,
  ``The strong decays of $K_1$ resonances,''
  Phys.\ Rev.\ D {\bf 85}, 074011 (2012)
  [arXiv:1111.6307 [hep-ph]].

\bibitem{kou3}
  E.~Kou, A.~Le Yaouanc and A.~Tayduganov,
  ``Angular analysis of $B \to J/\psi K_1$ : towards a model independent determination of the photon polarization with $B\to K_1$ gamma,''
  Phys.\ Lett.\ B {\bf 763}, 66 (2016)
  [arXiv:1604.07708 [hep-ph]].
 
\bibitem{gronau3}
   M.~Gronau and D.~Pirjol,
  ``Reexamining the photon polarization in $B\to K\pi\pi\gamma$,''
  Phys.\ Rev.\ D {\bf 96}, no. 1, 013002 (2017)
  [arXiv:1704.05280 [hep-ph]].
  
\bibitem{Dumm:2009va} 
  D.~G.~Dumm, P.~Roig, A.~Pich and J.~Portoles,
  ``tau ---> pi pi pi nu(tau) decays and the a(1)(1260) off-shell width revisited,''
  Phys.\ Lett.\ B {\bf 685}, 158 (2010)
  [arXiv:0911.4436 [hep-ph]].
\bibitem{Shekhovtsova:2012ra} 
  O.~Shekhovtsova, T.~Przedzinski, P.~Roig and Z.~Was,
  ``Resonance chiral Lagrangian currents and $\tau$ decay Monte Carlo,''
  Phys.\ Rev.\ D {\bf 86}, 113008 (2012)
  [arXiv:1203.3955 [hep-ph]].
\bibitem{Nugent:2013hxa} 
  I.~M.~Nugent, T.~Przedzinski, P.~Roig, O.~Shekhovtsova and Z.~Was,
  ``Resonance chiral Lagrangian currents and experimental data for $\tau^-\to\pi^{-}\pi^{-}\pi^{+}\nu_{\tau}$,''
  Phys.\ Rev.\ D {\bf 88}, 093012 (2013)
  [arXiv:1310.1053 [hep-ph]].
  
  
  
    \bibitem{ks}
   J.~H.~K\"uhn and A.~Santamaria,
  ``Tau decays to pions,''
  Z.\ Phys.\ C {\bf 48}, 445 (1990).
  
  
  
  \bibitem{Ackerstaff:1997dv}
   K.~Ackerstaff {\it et al.} [OPAL Collaboration],
  ``A Measurement of the hadronic decay current and the tau-neutrino helicity in tau- $\to$ pi- pi- pi+ tau-neutrino,''
  Z.\ Phys.\ C {\bf 75}, 593 (1997).
 
 
 
 


\bibitem{Asner:1999kj} 
  D.~M.~Asner {\it et al.} [CLEO Collaboration],
  ``Hadronic structure in the decay tau- $\to$ tau-neutrino pi- pi0 pi0 and the sign of the tau-neutrino helicity,''
  Phys.\ Rev.\ D {\bf 61}, 012002 (2000)
  [hep-ex/9902022].
 




  
\bibitem{bhm}
   B.~K.~Bullock, K.~Hagiwara and A.~D.~Martin,
  ``Tau polarization and its correlations as a probe of new physics,''
  Nucl.\ Phys.\ B {\bf 395}, 499 (1993).
 
 
\bibitem{Fischer:1979fh} 
  R.~Fischer, J.~Wess and F.~Wagner,
  ``Decays of the Heavy Lepton $\tau$ and Chiral Dynamics,''
  Z.\ Phys.\ C {\bf 3}, 313 (1979).
 
\bibitem{mg} 
  J.~Alwall, M.~Herquet, F.~Maltoni, O.~Mattelaer and T.~Stelzer,
  ``MadGraph 5 : Going Beyond,''
  JHEP {\bf 1106}, 128 (2011)
  [arXiv:1106.0522 [hep-ph]]. 

\bibitem{taudecay} 
  K.~Hagiwara, T.~Li, K.~Mawatari and J.~Nakamura,
  ``TauDecay: a library to simulate polarized tau decays via FeynRules and MadGraph5,''
  Eur.\ Phys.\ J.\ C {\bf 73}, 2489 (2013)
  [arXiv:1212.6247 [hep-ph]].


\bibitem{pythia} 
  T.~Sjostrand, S.~Mrenna and P.~Z.~Skands,
  ``A Brief Introduction to PYTHIA 8.1,''
  Comput.\ Phys.\ Commun.\  {\bf 178}, 852 (2008)
  [arXiv:0710.3820 [hep-ph]].

 
 \bibitem{mlm} 
  M.~L.~Mangano, M.~Moretti and R.~Pittau,
  ``Multijet matrix elements and shower evolution in hadronic collisions: $W b \bar{b}$ + $n$ jets as a case study,''
  Nucl.\ Phys.\ B {\bf 632}, 343 (2002)
  [hep-ph/0108069].
  
 \bibitem{delphes} 
  J.~de Favereau {\it et al.} [DELPHES 3 Collaboration],
  ``DELPHES 3, A modular framework for fast simulation of a generic collider experiment,''
  JHEP {\bf 1402}, 057 (2014)
  [arXiv:1307.6346 [hep-ex]].


 
 

 \bibitem{wigner}
   E.~P.~Wigner,
  ``On Unitary Representations of the Inhomogeneous Lorentz Group,''
  Annals Math.\  {\bf 40}, 149 (1939)
  [Nucl.\ Phys.\ Proc.\ Suppl.\  {\bf 6}, 9 (1989)].

  
 
 
\end{thebibliography}
\end{document}